\documentclass[11pt]{article}

\usepackage{graphicx}
\usepackage{latexsym}

\usepackage{amssymb,amsmath}
\usepackage{epstopdf}
\usepackage{color}
\normalbaselines

\catcode `\@=11
\@addtoreset{equation}{section}

\catcode`\@=12

\newtheorem{theo}{Theorem}[section]
\newtheorem{lem}[theo]{Lemma}
\newtheorem{prop}[theo]{Proposition}
\newtheorem{cor}[theo]{Corollary}

 \newtheorem{definition}[theo]{Definition}
\newtheorem{remark}[theo]{Remark}
\newtheorem{example}[theo]{Example}
 \numberwithin{equation}{section}

\newcommand{\betheo}{\begin{theo}}
\newcommand{\entheo}{\end{theo}}

\newcommand{\becor}{\begin{cor} }
\newcommand{\encor}{\end{cor}}

\newcommand{\belem}{\begin{lem}}
\newcommand{\enlem}{\end{lem}}

\newcommand{\beprop}{\begin{prop} }
\newcommand{\enprop}{\end{prop}}

\newcommand{\bedefi}{\begin{definition}$\!\!\!$ \rm }
\newcommand{\findefi}{ \end{definition}}

\newcommand{\beex}{\begin{example}$\!\!\!$ \rm }
\newcommand{\enex}{ \end{example}}

\newcommand{\berem}{\begin{remark}$\!\!\!$ \rm }
\newcommand{\enrem}{ \end{remark}}

\newcommand{\prf}{\noindent{\bf Proof.\,\,\,}}
\newcommand{\qed}{\hfill $\square$}

\newcommand{\be}{\begin{equation}}
\newcommand{\en}{\end{equation}}

\newcommand{\bea}{\begin{eqnarray}}
\newcommand{\ena}{\end{eqnarray}}

\newcommand{\beano}{\begin{eqnarray*}}
\newcommand{\enano}{\end{eqnarray*}}

\newcommand{\bee}{\begin{enumerate}}
\newcommand{\ene}{\end{enumerate}}

\newcommand{\bei}{\begin{itemize}}
\newcommand{\eni}{\end{itemize}}

\newcommand{\ta}{^\times}

\renewcommand{\leq}{\leqslant}
\renewcommand{\geq}{\geqslant}

 \newcommand{\noi}{\noindent}

 \newcommand{\ov}{\overline}
\newcommand{\ud}{\,\mathrm{d}}
\newcommand{\hs}{Hilbert space}
\newcommand{\pip}{{\sc pip}-space}
\newcommand{\ipip}{indexed {\sc pip}-space}
\newcommand{\spip}{{\sc pip}-subspace}
 \newcommand{\co}{^{\#}}

\newcommand{\com}{{\scriptstyle\#}}
\newcommand{\w}{\,\scriptstyle \ast\, }

\newcommand{\RN}{\mathbb{R}}

\newcommand{\BN}{{\mathbb N}}
\newcommand{\CN}{{\mathbb C}}
\newcommand{\ip}[2]{\langle {#1}|{#2}\rangle}

\def\A{{\mathcal A}}
\def\B{{\mathcal B}}
\def\C{{\mathcal C}}
\def\D{{\mathcal D}}
\def\E{{\mathcal E}}
\def\F{{\mathcal F}}
\def\H{{\mathcal H}}
\def\I{{\mathcal I}}
\def\cS{{\mathcal S}}

\def\gA{{\mathfrak A}}

\begin{document}

\begin{center}
{\LARGE Partial inner product spaces: Some categorical aspects} \vspace*{10mm}

{\large   J-P. Antoine$^{\rm a}$\footnote{{\it E-mail address}: jean-pierre.antoine@uclouvain.be},  
D. Lambert$^{\rm c}$\footnote{{\it E-mail address}: dominique.lambert@fundp.ac.be} and
C. Trapani$^{\rm b}$\footnote{{\it E-mail address}: trapani@unipa.it}
}
\\[3mm]
$^{\rm a}$ \emph{\small Institut de Recherche en Math\'ematique et  Physique, Universit\'e catholique de Louvain \\
B-1348   Louvain-la-Neuve, Belgium}
\\
$^{\rm b}$ \emph{\small Facult\'es Universitaires Notre-Dame de la Paix\\
 B-5000 Namur, Belgium }
 \\
$^{\rm c}$ \emph{\small Dipartimento di Matematica e Informatica,Universit\`a di Palermo\\
I-90123 Palermo, Italy }
\end{center}

\begin{abstract} \noindent
We make explicit in terms of categories a number of statements from the theory of partial inner product spaces (\pip s) and operators on them.
 In particular, we   construct sheaves and cosheaves of operators on certain \pip s of practical interest.

\end{abstract}

\section{Motivation}

Partial inner product spaces (\pip s) were introduced some time ago by A. Grossmann and one of us (JPA) as a structure unifying
many constructions introduced in functional analysis, such as distributions or generalized functions, scales of Hilbert or Banach spaces, interpolation couples, etc. \cite{pip1}-\cite{pip4}. 
Since these structures have regained a new interest in many aspects of mathematical physics and in modern signal processing, 
a comprehensive monograph was recently published by two of us   \cite{pip-book}, as well as a review paper \cite{at-AMP}.

Roughly speaking, a \pip\ is a vector space equipped with a partial inner product, that is, an inner product which is not defined everywhere, but only for specific pairs of vectors. 
Given such an object, operators can be defined on it, that generalize the familiar notions of operators on a \hs, while admitting extremely singular ones.

Now, in the previous work, many statements have a categorical ``flavor", but   the corresponding
technical language was not used, only some hints in that direction were given   in \cite{iwota}. Here we fill the gap and proceed systematically.
We introduce the category ${\bf PIP}$ of (indexed) \pip s, with homomorphisms as arrows (they are defined precisely to play that role),   
as well as several other categories of \pip s.

In a second part, we consider a single \pip\ $V_{I}$ as a category by itself, called \textbf {V\boldmath$\!_I$}, with natural embeddings as arrows. 
For this category \textbf {V\boldmath$\!_I$}, we show, in Sections \ref{sec-sheaves} and  \ref{sec-cosheaves}, respectively,
that one can construct sheaves and cosheaves of operators. 
There are some restrictions on the \pip\ $V_{I}$, but the cases covered  by our results are the most useful ones for applications.
Then, in Section \ref{sec-cohomol}, we describe the cohomology of these (co)sheaves and prove that, in many cases, the sheaves of operators are acyclic, that is, all cohomology groups 
of higher order are trivial.

Although sheaves are quite common in many areas of mathematics, the same cannot be said of cosheaves, the dual concept of sheaves, for which very few concrete examples are known. 
Actually, cosheaves were recently introduced in the context of nonclassical logic (see Section \ref{sec-outcome})  and seem to be related to certain aspects of quantum gravity.
 Hence the interest of having at one's disposal new, concrete examples of cosheaves, namely,
cosheaves of operators on certain types of \pip s.

\section{Preliminaries}

\subsection{Partial inner product spaces}
 
We begin by fixing the terminology and notations, following our monograph \cite{pip-book}, to which we refer for a full information.
For the convenience of the reader, we have collected in   Appendix A the main features of partial inner product spaces and operators on them. 

Throughout the paper, we will consider an \ipip\  $V_I = \{V_r, r\in I\}$, corresponding to the linear compatibility $\com$. 
The assaying subspaces are denoted by $V_p, V_r,\ldots, p,r\in I$.
The set $I$ indexes a generating involutive sublattice of the complete lattice ${\F}(V,\com)$ of all assaying subspaces
defined by  $\com$, that is,
\be\label{eq:gener1}
f\com g \; \Leftrightarrow \; \exists \;r \in I \mbox{ such that } f \in V_{r}, g \in V_{\overline{r}}\,.
\en
The lattice properties are the following:
\smallskip

\begin{tabular}{lccl}
. involution: &$V_r$                     & $\!\!\!\leftrightarrow\!\!\!$& $V_{\ov{r}}:=(V_r)^{\#},$\\
. infimum:   & $V_{p \wedge q}$  &$\!\!\! :=\!\!\!$                & $V_p \wedge V_q = V_p \cap V_q,$   \qquad $(p,q,r \in I)$\\
. supremum: & $V_{p \vee q}$     &$\!\!\! :=\!\!\!$               &$ V_p \vee V_q = (V_p + V_q)^{\#\#}$.
\end{tabular}
\\[2mm]
The smallest element  of $ {\F }(V,\com)$ is  $V\co = \bigcap_r V_r $ and the greatest element   is $V = \bigcup_r V_r$, but often they do not belong to $V_I $.  

Each assaying subspace $V_r$ carries its Mackey topology  $\tau(V_{r},V_{\ov{r}})$ and $V\co$ is dense in every $V_r$, since the \ipip\ $V_I$ is assumed to be nondegenerate.
In particular, we consider   projective and additive \ipip s (see Appendix A) and, in particular, lattices of Banach or Hilbert spaces (LBS/LHS).

Given two \ipip s $V_I, Y_{K}$, an operator $A: V_I \to Y_{K}$ may be identified with the coherent collection of its representatives 
$A \simeq \{ A_{ur} \}$, where each $A_{ur}: V_{r} \to Y_{u}$ is a continuous operator from $V_{r}$ into $Y_{u}$. We will also need the set
${\sf d}(A)  = \{ r \in I : \mbox{there is a } \,   u\in K \; \mbox{such that}\; A_{ur} \;\mbox{exists} \}.$
Every operator $A$ has an adjoint $A\ta$ and a  partial multiplication between operators is defined.

A crucial role is played by homomorphisms, in particular, mono-, epi- and isomorphisms. The set of all operators from $V_I $ into $Y_{K}$ 
is denoted by $\mathrm{Op}(V_{I},Y_{K})$ and the set of all homomorphisms by  $\mathrm{Hom}(V_{I},Y_{K})$.

For more details   and references to related work, see   Appendix A or our monograph \cite{pip-book}.

\subsection{Categories}

According to the standard terminology \cite{maclane1}, a (small) category {\bf C} is a collection of objects $X,Y,Z,\ldots$ and arrows or morphisms $\alpha, \beta,\ldots$, where 
we note $\alpha \in \mathrm{hom}(X,Y)$ or $X \stackrel{\alpha}{\longrightarrow}Y$, satisfying the following axioms.
\bei
\vspace*{-2mm}\item {\sl Identity:} for any object $X$, there is a unique arrow $1_{X}: X\to X$.

\vspace*{-2mm}\item {\sl Composition:} whenever $X \stackrel{\alpha}{\longrightarrow}Y\stackrel{\beta}{\longrightarrow}Z $, there is a unique arrow
 $\beta \circ \alpha $ such that  $X \stackrel{\beta \circ\alpha}{\longrightarrow}Z$.

\vspace*{-2mm}\item {\sl Associativity:} whenever $X \stackrel{\alpha}{\longrightarrow}Y\stackrel{\beta}{\longrightarrow}Z \stackrel{\gamma}{\longrightarrow}T$,
one has $(\gamma \circ \beta) \circ \alpha =  \gamma \circ (\beta \circ \alpha)$.

\vspace*{-2mm}\item {\sl Unit law:} whenever $X \stackrel{\alpha}{\longrightarrow}Y$, one has $\alpha \circ 1_{X}=  \alpha $ and $1_{Y} \circ \alpha  = \alpha $.
\eni

In a category, an object $S$ is \emph{initial} if, for each object $X$, there is exactly one arrow $S \to X$.
An object $T$ is \emph{final} or \emph{terminal} if for each object $X$, there is exactly one arrow $X\to T$. Two terminal objects are necessarily isomorphic 
(isomorphisms in categories are defined exactly as for \ipip s, see   Appendix A).

Given two categories \textbf{C} and \textbf{D}, a \emph{covariant functor} ${\sf F}: \textbf{C}\to \textbf{D}$ is a morphism between the two categories.
 To each object $X$ of \textbf{C}, it associates an object ${\sf F}(X)$ of \textbf{D} and to each arrow $\alpha : X \to Y$ in \textbf{C}, it associates an arrow
${\sf F}(\alpha) : {\sf F}(X) \to {\sf F}(Y)$ of \textbf{D}  in such a way that
$$
{\sf F}(1_X) = 1_{{\sf F}(X)} \qquad \textrm{and} \qquad {\sf F}(\beta \circ \alpha) = {\sf F}(\beta) \circ {\sf F}(\alpha),
$$ 
whenever the arrow $\beta \circ \alpha$ is defined in \textbf{C}.

Given a category \textbf{C}, the \emph{opposite} category $\textbf{C}^{\rm op}$ has the same objects as  \textbf{C} and all arrows reversed: to each arrow $\alpha: X\to Y$, 
there is an arrow $\alpha^{\rm op}: Y\to X$, so that $\alpha^{\rm op} \circ \beta^{\rm op} = (\beta \circ \alpha)^{\rm op}$.

A \emph{contravariant functor} $\ov{\sf F}: \textbf{C}\to \textbf{D}$  may be defined as a functor ${\sf F}: \textbf{C}^{\rm op} \to \textbf{D}$, or directly on
\textbf{C}, by writing $\ov{\sf F}(\alpha) = {\sf F}(\alpha^{\rm op})$. Thus we have
$$
\ov{\sf F}(1_X) = 1_{\ov{\sf F}(X)} \qquad \textrm{and} \qquad \ov{\sf F}(\beta \circ \alpha) = \ov{\sf F}(\alpha) \circ \ov{\sf F}(\beta).
$$

Some standard examples of categories are

. \textbf{Set}, the category of sets with functions as arrows.

. \textbf{Top}, the category of topological spaces with continuous functions as arrows.

. \textbf{Grp}, the category of groups with group homomorphisms as arrows.

\noi For more details, we refer to standard texts, such as Mac Lane \cite{maclane1}.

\section{Categories of \pip s}
\subsection{A single \pip\ as category}
\label{subsec:single}
 We begin by a trivial example.  

A single \ipip\ $V_I = \{V_r, r\in I\}$ may be considered as a category \textbf {V\boldmath$\!_I$}, where
\bei
\vspace*{-2mm}\item The objects are the assaying subspaces $ \{V_r, r\in I, V\co, V\}$. 
\vspace*{-2mm}\item The   arrows are the natural embeddings $ \{E_{rs}: V_r \to V_s, \; r\leq s\}$, that is, the representatives of the identity operator on $V_{I}$. 
\eni
The axioms of categories are readily checked
\bei
\vspace*{-2mm}\item For every $V_r,$ there exists an identity, $ E_{rr} : V_r \to V_r$, the identity map.
\vspace*{-2mm}\item   
For every $V_r, V_s$ with $r\leq s$,  one has $ E_{ss} \circ E_{sr} = E_{sr} $ and $E_{sr}  \circ E_{rr}  =  E_{sr} $.
\vspace*{-2mm}\item   
For every $V_r, V_s, V_t$ with $r\leq s\leq t$, one has $ E_{ts}  \circ E_{sr} = E_{tr} $.
\vspace*{-2mm}\item   
For every $V_r, V_s, V_t, V_u$ with $r\leq s\leq t\leq u$,  one has $ (E_{ut}  \circ E_{ts} ) \circ E_{sr} = E_{ut} \circ (E_{ts} \circ E_{sr})$.
\eni
 
\noi
In the category \textbf {V\boldmath$\!_I$},
\vspace*{1mm}

\noi .  $V_{\ov\infty}:=V\co  = \bigcap_{r\in I}$ is an initial object: for every $\ V_r \in V_I,$   there is a unique arrow  $ E_{r\ov\infty}: V\co \to V_r  $.
\\[1mm]
.  $V_{\infty}:= V  = \sum_{r\in I}$ is a terminal  object: for every  $ V_r \in V_I, $ there is a unique arrow $ E_{\infty r}: V_r \to V $.
\\[1mm]
. The compatibility $\com : V_r \mapsto (V_r)\co = V_{\ov r} $ defines a contravariant functor $\textbf {V\boldmath$\!_I$}\to \textbf {V\boldmath$\!_I$}$.
\vspace*{1mm}

Although this category seems rather trivial, it will allow us to define sheaves and cosheaves of operators,  a highly nontrivial (and desirable) result.

\subsection{A  category generated by a single operator}

  In the \ipip\ $V_I = \{V_r, r\in I\}$, take a single totally regular operator, that is, an operator $A$ that leaves every $V_r$  invariant. Hence so does each power $A^n, \,n\in \BN$. 
Then this operator induces a category  \textbf {A({V\boldmath$\!_I$})}, as follows:
\bei
\vspace*{-2mm}\item   
The objects are the assaying subspaces $ \{V_r, r\in I\}$.
\vspace*{-2mm}\item The arrows are the operators  $ A^n_{pq}: V_q \to V_p, \, q\leq p, \, n\in \BN$. 
\eni
 The axioms of categories are readily checked:
\bei
\vspace*{-2mm}\item For every $ V_r$, there exists an identity, $A_{rr} : V_r \to V_r  $, since $A$ is totally regular. 
\vspace*{-2mm}\item   
For every $ V_r, V_s $ with $r\leq s$, one has $ A_{ss} \circ A^n_{sr} = A^{n+1}_{sr} $ and $A^n_{sr}  \circ A_{rr}  =  A^{n+1}_{sr},\,  \forall\, n\in \BN$.
\vspace*{-2mm}\item   
For every $V_r, V_s, V_t$ with $r\leq s\leq t$, one has $ A^m_{ts}  \circ A^n_{sr} = A^{m+n}_{tr},\,  \forall\,m, n\in \BN $.
\vspace*{-2mm}\item   
For every $V_r, V_s, V_t, V_u$ with $r\leq s\leq t\leq u$,  one has $ (A^m_{ut}  \circ A^n_{ts} ) \circ A^k_{sr} = A^m_{ut} \circ (A^n_{ts} \circ A^k_{sr}),\,  \forall\,m, n,k\in \BN $.
\eni
 As for  \textbf {V\boldmath$\!_I$},  the space $V_{\ov\infty}:= V\co$ is an initial object in \textbf {A({V\boldmath$\!_I$})} and
 $V_{\infty}:= V$ is a terminal object.

  The adjunction $A\mapsto A\co$ defines a contravariant functor from \textbf {A({V\boldmath$\!_I$})} into \textbf {A({V\boldmath$\!_I$})}$\co$, 
where the latter is the category induced by $A\co$.
The proof is immediate.

\subsection{\color{black}The category ${\bf PIP}$ of \ipip s}
\label{subsec:PIP}

The collection of all  \ipip s constitutes a category, that we call  {\bf PIP}, where
\bei
\vspace*{-2mm}\item Objects are \ipip s $ \{V_I\}$. 
\vspace*{-2mm}\item Arrows are homomorphisms  $A: V_I \to Y_K$,   
where an operator     $A\in \mathrm{Op}(V_{I},Y_{K})$ is called a \emph{homomorphism} if
\bei
\vspace*{-2mm}\item [(i)] for every $r\in I$ there exists $u\in K$ such that both $A_{ur}$ and $A_{\ov u \ov r}$  exist;
\vspace*{-2mm}\item [(ii)] for every $u\in K$ there exists $r\in I$ such that both $A_{ur}$ and $A_{\ov u \ov r}$  exist. 
\eni

\eni
For making the notation less cumbersome (and more automatic), we will henceforth denote by $A_{KI}$ an element $A \in \mathrm{Hom}(V_{I},Y_{K})$.
Then the axioms of a category are obviously satisfied:
\bei
\vspace*{-2mm}\item For every  $V_I$, there exists an identity, $1_I \in \mathrm{Hom}(V_{I},V_{I})$,  the identity operator on $V_{I}$.
\vspace*{-2mm}\item For every $V_I, Y_K, $ one has $\; 1_K \circ A_{KI} = A_{KI}$ and $A_{KI} \circ \; 1_I =  A_{KI}$.
\vspace*{-2mm}\item For every $V_I, Y_K, W_L,$ one has $ B_{LK} \circ A_{KI}= C_{LI} \in \mathrm{Hom}(V_{I},W_{L})$.
\vspace*{-2mm}\item For every $V_I, Y_K, W_L, Z_M$,  one has 
  
{$ (C_{ML} \circ B_{LK}) \circ A_{KI}= C_{ML} \circ (B_{LK} \circ A_{KI})\in \mathrm{Hom}(V_{I},Z_M)$.}
\eni
The category ${\bf PIP}$ has no initial object and no terminal object, hence it is not a topos.
\vspace*{2mm} 

One can define in the same way smaller categories {\bf LBS} and {\bf LHS}, whose objects are, respectively,  lattices of Banach spaces (LBS) 
and lattices of \hs s (LHS), the arrows being still the corresponding homomorphisms.

\subsubsection{Subobjects}
 \label{subsubsec:subobjects}

 We recall that a homomorphism $M_{KI} \in \mathrm{Hom}(V_{I},Y_{K})$ is a monomorphism if $M_{KL}A_{LI}=M_{KL}B_{LI}$ implies $A_{LI}=B_{LI}$,
 for any pair $A_{LI},B_{LI}$ and any \ipip\ $W_L$ (a typical example is given in Section \ref{sec:oper}). Two monomorphisms  $M_{LI}, N_{LK}$ with the same codomain $W_L$ are \emph{equivalent} if there exists an isomorphism
 $U_{KI}$ such that $N_{LK}U_{KI}= M_{LI}$. Then a \emph{subobject} of $V_I$ is an equivalence class of monomorphisms into $V_I$.
A  \spip\ $W$ of an \ipip\ $V$ is defined as an orthocomplemented subspace of $V$ and   this holds
 if and only if $W$  is the range of an orthogonal projection,  $W_I=PV_I$.   
 Now the embedding $M: W_I=PV_I\mapsto V_I$ is a monomorphism, thus  orthocomplemented subspaces are subobjects of  ${\bf PIP}$. 

  However, the converse is not true,    at least for a general \ipip.  Take the case where V is a non-complete prehilbert space (\emph{i.e.},  $V=V\co$). 
Then every subspace is a subobject, but  need not be the range of a projection.
To give a concrete example \cite[Sec. 3.4.5] {pip-book}, take $V ={\mathcal S}(\RN)$, the Schwartz space of test functions.
Let $W ={\mathcal S}_{+}:= \{\varphi \in {\mathcal S} :\varphi(x) = 0 \; \mathrm{for} \; x\leq 0 \}$. 
Then $W^\bot ={\mathcal S}_{-}:= \{\psi \in {\mathcal S} : \psi(x) = 0 \;  \mathrm{for} \; x\geq 0 \}$, 
hence $W^{\bot \bot} = W$. However, $W$ is not orthocomplemented, since every $\chi \in W + W^\bot$ satisfies $\chi (0) = 0$, so that $W + W^\bot \neq {\mathcal S}$. 
Yet $W$ is the range of a monomorphism (the injection), hence a subobject. However this example addresses an \ipip\ which is \emph{not} a LBS/LHS. 

 Take now a LBS/LHS $V_I = \{V_r, r\in I\}$ and a vector subspace $W$. In order that $W$ becomes a  LBS/LHS $W_{I}$ in its own right, we must require that, for every $r\in I$,
$W_{r} = W \cap V_{r}$  and $W_{\ov{r}} = W\cap V_{\ov{r}} $ are a dual pair with respect to their respective Mackey topologies and that
 the intrinsic Mackey topology $\tau (W_{r}, W_{\ov{r}})$ coincides with the norm topology   induced by $V_{r}$. In other words, $W$ must be topologically regular,
which is equivalent that it be orthocomplemented \cite[Sec.3.4.2]{pip-book}. Now the injection $M_{I}: W_{I}\to V_{I}$ is clearly a monomorphism and $W_{I}$ is a subobject of $V_{I}$. 
Thus we have shown that, in a LBS/LHS, the subobjects are precidely the orthocomplemented subspaces.

Coming back to the previous example of a non-complete prehilbert space, we see that an arbitrary subspace $W$ need \emph{not} be orthocomplemented, because it may fail to be
topologically regular. 
 Indeed the intrinsic topology $\tau (W, W)$ does not coincide with the norm topology, unless $W$ is orthocomplemented
(see the discussion in \cite[Sec. 3.4.5] {pip-book}).
In the Schwartz example above, one has $W^{\bot \bot} = W$, which means that $W$ is $\tau (W, W)$-closed, hence norm-closed, but it is not  orthocomplemented.

   \berem
Homomorphisms are defined between arbitrary \pip s. However, when it comes to \emph{indexed} \pip s,
the discussion above shows that the notion of homomorphism is more natural between two \ipip s \emph{of the same type}, for instance, two LBSs or two LHSs.
This is true, in particular, when trying to identify subobjects.
This suggests to define   categories {\bf LBS} and {\bf LHS}, either directly as above, or as subcategories within {\bf PIP}, and then define properly subobjects in that context.
\enrem
  
\subsubsection{Superobjects}

Dually, one may define  superobjects in terms of  epimorphisms.
We recall that a homomorphism $N_{KL} \in \mathrm{Hom}(V_{I},W_{K})$ is an epimorphism if $A_{IK}N_{KL}=B_{IK}N_{KL}$ implies $A_{IK}=B_{IK}$,
 for any pair $A_{IK},B_{IK}$  and any \ipip\ $Y_L$. Then a  \emph{superobject} is an equivalence class of epimorphisms, where again equivalence means modulo isomorphisms.
 
Whereas monomorphisms are natural in the context of sheaves, epimorphisms  are natural in the dual structure, \emph{i.e.}, cosheaves.

\section{Sheaves  of operators on \pip s}
\label{sec-sheaves}

\subsection{Presheaves and sheaves}

Let $X$ be a topological space, and let ${\bf C}$ be a (concrete) category. Usually ${\bf C}$ is the category of sets, the category of groups, the category of abelian groups, 
or the category of commutative rings.   In the standard fashion \cite{maclane1,maclane2}, we proceed in two steps.  

\medskip
\bedefi \label{defn_presheaf} A \emph{presheaf} ${\sf F}$ on $X$ with values in ${\bf C}$ is a map ${\sf F}$ defined on the family of open subsets of $X$ such that:
\begin{itemize}
\item[(PS$_1$)] for each open set $U$ of $X$,    there is an object ${\sf F}(U)$ in ${\bf C}$;
\item[(PS$_2$)]  
 for each inclusion of open sets $T \subseteq U$, there  is given a \emph{restriction morphism}  $\rho_V^T : {\sf F}(U)\to {\sf F}(T)$ in the category ${\bf C}$, 
such that $\rho_U^U$ is the identity for every open set $U$  and $\rho_S^T \circ \rho_T^U = \rho_S^U$ whenever $S \subseteq T \subseteq U$.
\end{itemize}

\findefi

\bedefi Let ${\sf F}$ be a presheaf on $X$ and $U$ an open set of $X$. Every element $s\in {\sf F}(U)$ is called a {\em section} of ${\sf F}$ over $U$. 
A section over $X$ is called a \emph{global section}.
\findefi

\beex Let $X$ be a topological space, ${\bf C}$ the category of vector spaces. Let ${\sf F}$ associate to each open set $U$ the vector space of continuous functions on 
$U$ with values in $\CN$.   
 If $T\subseteq U$, $\rho_T^U$ associates to each continuous function on $U$ its restriction to $T$.  This is a presheaf.
Any continuous function on $U$ is a section of ${\sf F}$ on $U$.
\enex

\bedefi 
\label{def-presheaf}
Let ${\sf F}$ be a presheaf on the topological space $X$. We say that ${\sf F}$ is a \emph{sheaf} if, for every open set $U\subset X$ and  
for every open covering $\{U_i\}_{i\in I}$ of $U$, 
the following conditions are fulfilled:
\begin{itemize}
\item[(S$_1$)] given $s, s' \in {\sf F}(U)$ such that $\rho^{U}_{U_i}(s)= \rho^{U}_{U_i}(s')$, for every $i\in I$, then $s=s'$ (local identity);
\item[(S$_2$)] given $s_i \in {\sf F}(U_i)$ such that $\rho^{U_i}_{U_i \cap U_j} (s_i)=\rho^{U_j}_{U_i \cap U_j} (s_j)$, for every $i,j \in I$, then there exists a section 
$s\in {\sf F}(U)$  such that $\rho^U_{U_i}(s)=s_i$, for every $i \in I$ (gluing).
\end{itemize}
\findefi
 
The section $s$ whose existence is guaranteed by axiom S$_2$ is called the {\em gluing}, {\em concatenation}, or {\em collation} of the sections $s_i$. By axiom S$_1$ it is unique.
The sheaf ${\sf F}$ may be seen as a contravariant functor from the category of open sets of $X$ into  
${\bf C}:= (\{{\sf F}(U_i)\}, \{\rho^{U_i}_{U_j})\}$.

 \subsection{A sheaf of operators on an \ipip}

Let $V_I = \{V_r, r\in I\}$ be an \ipip\ and \textbf { V$_I$} the corresponding category defined in Section \ref{subsec:single}. 
If we put on $I$ the discrete topology, then $I$ defines an open covering of $V$.
Each $V_r$ carries its Mackey topology $\tau (V_{r}, V_{\ov{r}})$.

We define a sheaf on $V_I$ by the contravariant functor ${\sf F} :{\bf V}_I \to {\bf Set}$ given by
\begin{align} \label{eq:opsheaf}
{\sf F}: V_r \mapsto  \mathrm{Op}_r  &:=\{A\upharpoonright V_r : A \in \mathrm{Op}(V_I), r\in {\sf d}(A) \}.  
\end{align}
This means that an element of $\mathrm{Op}_r$ is a representative $A_{w r}$ from $V_r$ into some $V_w$.
In the sequel, we will use the notation  $A_{\w r}$    
 whenever the dependence on the first index may be neglected without creating ambiguities.
By analogy with functions, the elements   of  Op$_r$ may be called  germs of operators. Op$_r$ is the restriction to $V_r$ of the set of left multipliers \cite[Sec.6.2.3]{pip-book}
$$
L_r:= \{A\in  \mathrm{Op}(V_{I}) : \exists \, w \; \text{  such that }   A_{wr} \; \mathrm{  exists}\} = \{A\in  \mathrm{Op}(V_{I}) :   r \in {\sf d}(A) \}.
$$
Then we have
\begin{itemize}
\item When $V_q \subseteq V_p$, define $\rho^p_q: \mathrm{Op}_p \to \mathrm{Op}_q$
 by $\rho^p_q(A_{\w p}):= A_{\w q}$,  for $A_{\w p}\in \mathrm{Op}_p$.
Clearly, $\rho_p^p =id_{\mathrm{Op}_p}$ and $\rho_r^q \circ \rho^p_q = \rho_r^p$ if $V_r \subset V_q \subset V_p$. Hence
${\sf F}$ is a presheaf.

\item (S1) is clearly satisfied. As for (S2), if $A_{\w r} \in \mathrm{Op}_r$ and $A'_{\w s} \in \mathrm{Op}_s$ are such that
$\rho^{r}_{r \wedge s} (A_{\w r})=\rho^{s}_{r \wedge s} (A'_{\w s})$,
that is, $ A_{\w r \wedge s} =A'_{\w r \wedge s}$, then these two operators are the 
 $(r \wedge s,\w)$-representative of a unique operator $A \in\mathrm{Op}(V_I)$. It remains to prove that $A$ extends to $V_{r \vee s}$ 
and that its representative $A_{\w r \vee s}$ extends both $A_{\w r}$ and $A'_{\w s}$. 
\end{itemize}
\beprop
Let the \ipip\ $V_I$ be additive, that is, $V_{r \vee s} = V_r + V_s, \forall\, r,s\in I$. Then the map ${\sf F}$ given in \eqref{eq:opsheaf}
is a sheaf of operators on $V_I$.
\enprop
\prf By linearity,   $A_{\w r}$ and $A_{\w s}$ can be extended to an operator $A^{(r\vee s)}$  on $V_r + V_s$, as follows:
$$
A^{(r\vee s)}(f_r +f_s):= A_{\w r}f_r +A_{\w s}f_s.
$$
This operator is well-defined. Let indeed $f=f_r +f_s= f'_r +f'_s$ be two decompositions of $f\in V_r + V_s$,
so that $f_r -f'_r = f'_s - f_s \in V_r \cap V_s$. Then
$A^{(r\vee s)}f = A_{\w r}f_r +A_{\w s}f_s = A_{\w r}f'_r +A_{\w s}f'_s$. Hence 
$A_{\w r}(f_r-f'_r)  = A_{\w s}(f'_s-f_s)$. Taking the restriction to $V_r \cap V_s= V_{r\wedge s}$, this relation becomes
$A_{\w  r\wedge s}(f-f')_{r\wedge s} = A_{\w  r\wedge s}(f'-f)_{r\wedge s}= 0$, so that the vector $A^{(r\vee s)}f$ is uniquely defined.

Next, by additivity, $V_r + V_s$, with its inductive topology, coincides with $V_{r \vee s}$ and thus $A_{\w r\vee s}$ is the 
$(r \vee s,\w)$-representative of   the  operator $A \in\mathrm{Op}(V_I)$. Therefore ${\sf F}$ is a sheaf.
\qed
\bigskip

We recall that most interesting classes of \ipip s are additive, namely, the projective ones and,   
 in particular,  LBSs and LHSs.  Thus the proposition just proven has a widely applicable range.

\section{Cosheaves  of operators on \pip s}

\label{sec-cosheaves}

\subsection{Pre-cosheaves and cosheaves}

Pre-cosheaves and cosheaves are the dual notions of presheaves and sheaves, respectively.
Let again $X$ be a topological space, with closed sets $W_i$ so that $X = \bigcup_{i\in I}W_i$, and let ${\bf C}$ be a (concrete) category. 
\medskip
\bedefi \label{defn_precosheaf} A \emph{pre-cosheaf} ${\sf G}$ on $X$ with values in ${\bf C}$ is a map ${\sf G}$ defined on the family of closed subsets of $X$ such that:
\begin{itemize}
\item[(PC$_1$)]   
for each closed set $W$ of $X$, there is an object ${\sf G}(W)$ in ${\bf C}$;
\item[(PC$_2$)] for each inclusion of closed sets $Z \supseteq W$, there  is given a \emph{extension morphism}  $\delta_W^Z : {\sf G}(W)\to {\sf G}(Z)$ in the category ${\bf C}$, 
such that $\delta_W^W$ is the identity for every closed set $W$ and $\delta_Z^T \circ \delta_W^Z = \delta_W^T$ whenever $T \supseteq Z \supseteq W$.

\end{itemize}
\findefi

\bedefi Let ${\sf G}$ be a pre-cosheaf on the topological space $X$. We say that ${\sf G}$ is a \emph{cosheaf} if, for every nonempty closed set $W=\bigcap_{j\in J}W_j, \, J \subseteq I$ 
and for   every family of (local) sections  $\{t_j \in {\sf G}(W_j)\}_{j\in J}$, the following conditions are fulfilled:
\begin{itemize}
\item[(CS$_1$)]  given $t,t' \in {\sf G}(W)$ such that  $\delta_{W}^{W_j}(t)= \delta_{W}^{W_j}(t')$, for every $j\in I$ such that $W\subseteq W_j$, then $t=t'$; 

\item[(CS$_2$)] if $\delta_{W_j}^{W_j \cup W_k} (t_j)=\delta_{W_k}^{W_j \cup W_k} (t_k)$, for every $j,k\in J$, then there exists a unique  section $t\in {\sf G}(W)$   such that $\delta_W^{W_j}(s)=t_j$, for every $j \in J$ .
\end{itemize}
\findefi

The cosheaf ${\sf G}$ may be seen as a covariant functor from the category of closed sets of $X$ into  
${\bf C}:= (\{{\sf G}(W_i)\}, \{\delta^{W_i}_{W_j})\}$.
\medskip

Now, there are situations where extensions do not exist for all inclusions $Z \supseteq W$, but only for certain pairs. This will be the case for operators on a \pip, as will be seen below. 
Thus we may generalize the notions of (pre)-cosheaf as follows (there could be several variants).

\bedefi \label{defn_pprecosheaf} Let  $X$  be a  topological space, ${\mathcal C l}(X)$ the family of closed subsets of $X$ and $\preceq$ a coarsening of the set inclusion in  
${\mathcal C l}(X)$, that is, $W\preceq Z$ implies $W\subseteq Z$, but not necessarily the opposite.
 A \emph{partial}\,\footnote{
 We may use the term `partial', since here not all pairs $w\leq z$ admit extensions, but only certain pairs, namely, those that satisfy $w\preceq z$. 
This is analogous to the familiar situation of a compatibility.}   \emph{pre-cosheaf}  ${\sf G}$ 
with values in ${\bf C}$ is a map ${\sf G}$ defined on ${\mathcal C l}(X)$, satisfying condition (PC$_1$) and
\begin{itemize}
\item[(pPC$_2$)] If $Z\succeq W$,  there  is given an \emph{extension morphism}  $\delta_W^Z : {\sf G}(W)\to {\sf G}(Z)$ in the category ${\bf C}$,  
 such that $\delta_W^W$ is the identity for every closed set $W$ and $\delta_Z^T \circ \delta_W^Z = \delta_W^T$ whenever $T \succeq Z \succeq W$.
 \end{itemize}
\findefi

\bedefi Let ${\sf G}$ be a partial pre-cosheaf on the topological space $X$. We say that ${\sf G}$ is a \emph{partial cosheaf} if, for every nonempty closed set $W=\bigcap_{j\in J}W_j, \, J \subseteq I$ 
and for  every family of sections  $\{t_j \in {\sf G}(W_j)\}_{j\in J}$,  the following conditions are fulfilled:
\begin{itemize}
\item[(pCS$_1$)] given $t,t' \in {\sf G}(W)$ such that  $\delta_{W}^{W_j}(t) $ and $\delta_{W}^{W_j}(t')$ exist and are equal, for every $j \in J$, then $t=t'$; 

\item[(pCS$_2$)] if $\delta_{W_j}^{W_j \cup W_k} (t_j) $ and $\delta_{W_k}^{W_j \cup W_k} (t_k)$ exist and are equal, for every $j,k \in J$, then 
$\delta_W^{W_j}$ exists for every $j \in J$ and
there exists a   unique section $t\in {\sf G}(W)$   such that $\delta_W^{W_j}(t)=t_j$, for every $j \in J$ .
\end{itemize}
\findefi

\subsection{Cosheaves of operators on \ipip s}

Let again $V_I = \{V_r, r\in I\}$ be an \ipip. If we put on $I$ the discrete topology, then the assaying subspaces may be taken as 
closed sets in $V$. In order to build a cosheaf, we consider the same map as before in \eqref{eq:opsheaf}:
\begin{align*}
{\sf F}: V_r \mapsto  \mathrm{Op}_r &:=  \{A_{wr} : V_r \to V_w, \mbox{ for some } w\in I, A \in \mathrm{Op}(V_I) \}.  
\end{align*}
Then we immediately face a problem. Given $V_q$, there is no $V_p$, with $V_q \subseteq V_p$, such that $\delta^p_q: \mathrm{Op}_p \to \mathrm{Op}_q$ exists. 
One can always find an operator $A$ such that $q\in {\sf d}(A)$ and $p\not\in {\sf d}(A)$, in other words, $A_{\w q}$ exists, but it cannot be extended to $V_p$. 
 This happens, for instance, each time ${\sf d}(A)$ has a maximal element (see Figure 3.1 in \cite{pip-book}). 
There are three ways out of this situation.

\subsubsection{General operators: the additive case}\label{sec521}

We still consider the set of all operators $\mathrm{Op}(V_I)$. Take two assaying subspaces $V_r,V_s$ and two operators
$A^{(r)} \in \mathrm{Op}_r, B^{(s)} \in \mathrm{Op}_s$ and assume that they have a common extension $C^{(r\vee s)}$ to $\mathrm{Op}_{r\vee s}$.
This means that, for any suitable $w$, $ C^{(r\vee s)}: V_{r \vee s} \to V_w$ is the $(r\vee s,w)$-representative $C_{w,r\vee s}$ of a unique operator $C\in \mathrm{Op}(V_I)$, and $C=A=B$.
\emph{A fortiori}, $A$ and $B$ coincide on $V_{r\wedge s}= V_r \cap V_s$, that is, condition (pCS$_2$) is satisfied.

Assume now that $V_I$ is additive, \emph{i.e.}, $V_{r \vee s}$ coincides with $V_r + V_s$, with its inductive topology. Then we can  proceed as in the case of a sheaf, by defining the extension $C$ of $A$ and $B$ by linearity. 
In other words, since $C^{(r\vee s)}f_r = A_{\w r}f_r$ and $C^{(r\vee s)}f_s= B_{\w s}f_s$, we 
may write, for any $f=f_r +f_s \in V_r + V_s$,
$$
C^{(r\vee s)}f = C_{\w r \vee s}f_{r\vee s}= A_{\w r}f_r +B_{\w s}f_s.
$$
Here too, this operator is  well-defined. Let indeed $f=f_r +f_s= f'_r +f'_s$ be two decompositions of $f\in V_r + V_s$,
so that $f_r -f'_r = f'_s - f_s \in V_r \cap V_s$. Then necessarily
$C^{(r\vee s)}f = A_{\w r}f_r +B_{\w s}f_s = A_{\w r}f'_r +B_{\w s}f'_s$. Hence 
$A_{\w r}(f_r-f'_r)  = B_{\w s}(f'_s-f_s)$. Taking the restriction to $V_r \cap V_s= V_{r\wedge s}$, this relation becomes
$A_{\w r\wedge s}(f-f')_{r\wedge s} = B_{\w   r\wedge s}(f'-f)_{r\wedge s}$, and this equals 0, since $A=B$ on $V_{r\wedge s}$, by 
 the condition (pcS$_2$).  

The conclusion is that, for any pair of assaying subspaces $V_r,V_s$, the extension maps $\delta_r^{r\vee s}$ and $\delta_s^{r\vee  s}$ always exist and the condition
 (pCS$_2$) is satisfied for that pair. 
 However, this is not necessarily true for any comparable pair and this motivates the coarsening of the order given in Proposition 
 \ref{prop410} below. Condition  (pPC$_2$) is also satisfied, as one can see by taking supremums (= sums) of  successive pairs within three assaying subspaces $V_r,V_s, V_t$ (and using the associativity of $\vee$). Thus me may state:
\beprop\label{prop410}
Let $V_I=   \{V_r, r\in I\}$ be an additive \ipip. Then the map ${\sf F}$ given in \eqref{eq:opsheaf} is a  partial cosheaf   with respect to the partial order on $I$
$$
r\preceq s \;\Leftrightarrow\; \exists\, w\in I \mbox{ such that } s=  r \vee w.
$$
\enprop

\subsubsection{Universal left multipliers}\label{sec522}

  We consider only the operators $A$ which are everywhere defined, that is, ${\sf d}(A)=I$. This is precisely the set of universal left multipliers
\begin{align*}
L \mathrm{Op}(V_{I}) & = \{C\in  \mathrm{Op}(V_{I}) : \forall \, r, \,\exists \,w \;\, \text{such that}  \; C_{wr} \; \mathrm{exists}\}\\
 &= \{ C \in  \mathrm{Op}(V_{I}) :   {\sf d}(C) =   I\}   
  =\textstyle\bigcap_{r \in I} L_r  , 
\end{align*} 
where $L_r:=  \{ C \in  \mathrm{Op}(V_{I}) :   r\in {\sf d}(C)\}$.
Correspondingly, we consider the map
\be \label{cosheafunivmult}
{\sf G}: V_r \mapsto  L\mathrm{Op}_r  :=\{A\upharpoonright V_r : A \in L \mathrm{Op}(V_I) \} . 
\en 
Here again, elements of $L\mathrm{Op}_r$ are representatives $A_{\w r}$.
In that case, extensions exist always. When $V_q \subseteq V_p$, define $\delta^p_q: L\mathrm{Op}_q \to L\mathrm{Op}_p$ 
by $\delta ^p_q(A_{\w q}):= A_{\w p} $ for $ A_{\w q}\in L \mathrm{Op}_q$.
Clearly, $\delta ^p _p=id_{L\mathrm{Op}_p}$ and $ \delta ^p_q \circ \delta ^q_r= \delta ^p_r$ if $V_r \subset V_q \subset V_p$. Hence
${\sf G}$ is a pre-cosheaf.
Moreover, if $ \delta ^{p\vee q}_q (A_{\w q}) =  \delta ^{p\vee q}_ p(A'_{\w p})$, that is,  $A_{\w,p\vee q} = A'_{\w,p\vee q}$, these two operators are the 
$(p\vee q,\w)$-representative of a unique operator $A\in L \mathrm{Op}(V_I)$. Then, of course, one has $ \delta _{p\wedge q}^q (A_{\wp\wedge q}) = A_{\w q}$ 
and $ \delta _{p\wedge q}^p (A_{\w \,p\wedge q}) = A_{\w p}$, thus ${\sf G}$ is a  cosheaf. Therefore,
\beprop
Let $V_I=   \{V_r, r\in I\}$ be an arbitrary \ipip. Then the map given in \eqref{cosheafunivmult} is a cosheaf on $V_I$ with values in 
 $L \mathrm{Op}(V_I)$, the set of universal left multipliers,  
with extensions given by $\delta ^p_q(A_{\w q})= A_{\w p}, \, q\leq p, \,A_{\w q}\in L \mathrm{Op}_q$. 
\enprop
 
\subsubsection{General operators: the projective case} \label{sec523}

We consider again all operators in  $ \mathrm{Op}(V_I)$. The fact is that  the set 
$\bigcup_{r\in I}\mathrm{Op}_r $  
is  an initial set in  $ \mathrm{Op}(V_I)$, like every {\sf d}(A), 
that is, it contains all predecessors of any of its elements, and this is natural for constructing sheaves into it. But it is \emph{not} when considering cosheaves. 
By duality one should rather take a final set, that is,  containing all successors of any of its elements, just like any {\sf i}(A). 
Hence we define the map:
\begin{align} \label{eq:opcosheaf} 
 \widehat{{\sf G}}: V_r \mapsto  \mathrm{\widehat {Op}}_r   :=\{A_{r\w} : A \in \mathrm{Op}(V_I), r\in {\sf i}(A) \}   
\end{align}
Now indeed   $\bigvee_{r\in I} \mathrm{\widehat {Op}}_r$ is a final set. 
Elements of $ \mathrm{\widehat {Op}}_r$ have been denoted as $A_{r\w}$, but, for definiteness, 
we could also replace $A_{r\w}$ by $A_{r{\ov\infty}}$   where $V_{\ov\infty}:=V\co$. 

We claim that the map $\widehat{\sf G}$ defines a cosheaf, with extensions $\delta ^p_q(A_{q\w }):= A_{p\w }, $ for $A_{q\w }\in \mathrm{\widehat {Op}}_q$,
which exist whenever $q\leq p$.

As in the case of Section \ref{sec521}, assume that  $A_{r\w } \in \mathrm{\widehat {Op}}_r, B_{s\w } \in \mathrm{\widehat {Op}}_s$ 
  have a common extension belonging 
to $\mathrm{\widehat {Op}}_{r\vee s}$, that is, $A_{r\vee s\w } := E_{r\vee s, r} A_{r\w } = B_{r\vee s \w }: = E_{r\vee s, s} B_{s\w }$. 
Call this operator $C_{r\vee s\w}$. Thus, for any suitable $w$,
$C_{r\vee s ,w}$ is the $(w,r\vee s )$-representative  of a unique operator $C\in \mathrm{Op}(V_I)$, 
which extends $A$ and $B$. Thus  $C=A=B \in \mathrm{Op}(V_I)$.
Assume now that $V_I$ is projective,  that is, $V_{r \wedge s} = V_r \cap V_s, \forall\, r,s\in I$, 
with the projective topology.  Then $C_{r\wedge s\w}=A_{r\wedge s\w}=B_{ r\wedge s\w}$, that is, condition (CS$_2$) is satisfied with extensions 
$\delta_{r\wedge s}^r (C_{r\wedge s \w } ) = A_{r\w } $ and  $\delta_{r\wedge s}^s (C_{r\wedge s \w } ) = B_{s\w } $.
The rest is obvious. As in the case of sheaves, extensions from $V_r,V_s$ to $V_{r\vee s}$ exist by linearity,  since $V_I$ is projective, hence additive. Thus we may state: 
 \beprop
Let the \ipip\ $V_I$ be projective, that is, $V_{r \wedge s} = V_r \cap V_s, \forall\, r,s\in I$, with the projective topology. 
Then the map $\widehat{\sf G}$ given in \eqref{eq:opcosheaf} is a cosheaf of operators on $V_I$.
\enprop

\becor\label{cor:co+sheaf}
Let $V_I=   \{V_r, r\in I\}$ be a projective \ipip. Then $V_I$ generates in a natural fashion a sheaf and a cosheaf of operators.
If $V_I$ is additive, but not projective, it generates  a sheaf and a partial cosheaf of operators.
\encor
 
\section{Cohomology of an  \ipip}
\label{sec-cohomol}

\subsection{Cohomology of operator sheaves} 
\label{sec-cohomolsheaf}

It is possible to introduce a concept of sheaf cohomology group defined on an \ipip\ according to the usual definition of sheaf cohomology \cite{neeman}. 
Let $V_I$ be an \ipip. The set of its assaying spaces, endowed with the discrete topology, defines an open covering of $V$. Endowed with the Mackey topology,  
$\tau(V_{r},V_{\ov{r}})$, each $V_{r}$  is a Hausdorff vector subspace of $V$.  

\bedefi \label{def_cochain}
Given the \ipip\ $V_I$, let $\{V_{j}, j\in I\}$ be the open covering of $V$ by its assaying subspaces   and let  
${\sf F}: V_r \mapsto  \mathrm{Op}_r =\{A\upharpoonright V_r : A \in \mathrm{Op}(V_I), r\in {\sf d}(A) \}$
be  the sheaf of operators defined in \eqref{eq:opsheaf} above.  We call \emph{$p$-cochain with values in ${\sf F}$} the map which associates to each intersection 
of open sets $V_{j_{0}} \cap V_{j_{1}}\cap \ldots\cap V_{j_{p}}, \, j_{k}\in I, \,j_{0} < j_{1}<\ldots < j_{p} $,
an operator  $A_{j_{0},j_{1},\ldots, j_{p}}$ belonging to  ${\sf F}(V_{j_{0}} \cap V_{j_{1}}\cap \ldots\cap V_{j_{p}})$. 
\findefi

Thus, a $p$-cochain with value in ${\sf F}$ is a set $\A= \{A_{j_{0},j_{1},\ldots, j_{p}}, \, j_{0} < j_{1}<\ldots < j_{p} \}$.
 The set of $p$-cochains is denoted by $C^p(I,{\sf F})$. For example, a 0-cochain is a set $\{A_{j}, j\in I\}$. A 1-cochain is a set  
$\{A_{j_{0},j_{1}}, \, j_{0} < j_{1} \}, \, A_{j_{0},j_{1}} \in {\sf F}(V_{j_{0}} \cap V_{j_{1}})$, \ldots
\bedefi \label{def_cobdy}
 Given the notion of $p$-cochain, the corresponding \emph{coboundary operator} $D$ is defined by
\begin{align*}
 &D: C^p(I,{\sf F}) \to C^{p+1}(I,{\sf F}): \{A_{j_{0},j_{1},\ldots, j_{p}}\}\mapsto \{(D\A)_{j_{0},j_{1},\ldots, j_{p},j_{p+1}}\},
\\
&(D\A)_{j_{0},j_{1},\ldots, j_{p},j_{p+1}}= 
\sum_{k=0}^{p+1}(-1)^{k}\rho_{V_{j_{0}} \cap V_{j_{1}}\cap \ldots\cap V_{j_{p+1}}}^{V_{j_{0}} \cap V_{j_{1}}\cap\ldots 
\cap \widehat V_{j_{k}}\cap\ldots\cap V_{j_{p+1}}}\, A_{j_{0},j_{1},\ldots, \widehat j_{k},\ldots,  j_{p+1}},
\end{align*}
where the hat corresponds to the omission of the corresponding symbol. We check that $D$ is a coboundary operator, \emph{i.e.}, $DD=0$. 
\findefi

If we compute the action of the coboundary operator on 0-cochains, we get
$$
(D\A)_{j_{0},j_{1}}  = \rho_{V_{j_{0}} \cap V_{j_{1}}}^{V_{j_{1}}} A_{j_{1}} - \rho_{V_{j_{0}} \cap V_{j_{1}}}^{V_{j_{0}}} A_{j_{0}}.
$$
This equation   is nothing but the condition (S$_{1}$) in Definition \ref{def-presheaf}, \emph{i.e.}, the necessary condition for getting  a sheaf.

We can also compute the action of  the coboundary operator on 1-cochains. We get:
$$
(D\A)_{j_{0},j_{1},j_{2}}  = \rho_{V_{j_{0}} \cap V_{j_{1}}\cap V_{j_{2}}}^{V_{j_{1}}\cap V_{j_{2}}} A_{j_{1},j_{2}}
 - \rho_{V_{j_{0}} \cap V_{j_{1}}\cap V_{j_{2}}}^{V_{j_{0}}\cap V_{j_{2}}} A_{j_{0},j_{2}}
+ \rho_{V_{j_{0}} \cap V_{j_{1}}\cap V_{j_{2}}}^{V_{j_{0}}\cap V_{j_{1}}} A_{j_{0},j_{1}}\, .
$$
Now, let us suppose that a 1-cochain is defined by $\A=D\B$, where $\B$ is a 0-cochain on $V$. The previous formula becomes:
\begin{align*}
(D\A)_{j_{0},j_{1},j_{2}}  = 
\rho_{V_{j_{0}} \cap V_{j_{1}}\cap V_{j_{2}}}^{V_{j_{1}}\cap V_{j_{2}}}
(\rho_{V_{j_{1}} \cap V_{j_{2}}}^{V_{j_{2}}} B_{j_{2}} - \rho_{V_{j_{1}} \cap V_{j_{2}}}^{V_{j_{1}}} B_{j_{1}})
&-
\rho_{V_{j_{0}} \cap V_{j_{1}}\cap V_{j_{2}}}^{V_{j_{0}}\cap V_{j_{2}}}
(\rho_{V_{j_{0}} \cap V_{j_{2}}}^{V_{j_{2}}} B_{j_{2}} - \rho_{V_{j_{0}} \cap V_{j_{2}}}^{V_{j_{0}}} B_{j_{0}})
\\
&+
\rho_{V_{j_{0}} \cap V_{j_{1}}\cap V_{j_{2}}}^{V_{j_{0}}\cap V_{j_{1}}}
(\rho_{V_{j_{0}} \cap V_{j_{1}}}^{V_{j_{1}}} B_{j_{1}} - \rho_{V_{j_{0}} \cap V_{j_{1}}}^{V_{j_{0}}} B_{j_{0}})
\end{align*}
Using the properties of restrictions we get:
\begin{align*}
(D\A)_{j_{0},j_{1},j_{2}}  = &\rho_{V_{j_{0}} \cap V_{j_{1}}\cap V_{j_{2}}}^{V_{j_{2}}}B_{j_{2}}  
-\rho_{V_{j_{0}} \cap V_{j_{1}}\cap V_{j_{2}}}^{V_{j_{1}}}B_{j_{1}}
-\rho_{V_{j_{0}} \cap V_{j_{1}}\cap V_{j_{2}}}^{V_{j_{2}}}B_{j_{2}}
\\ 
&+\rho_{V_{j_{0}} \cap V_{j_{1}}\cap V_{j_{2}}}^{V_{j_{0}}}B_{j_{0}} 
+\rho_{V_{j_{0}} \cap V_{j_{1}}\cap V_{j_{2}}}^{V_{j_{1}}}B_{j_{1}}
-\rho_{V_{j_{0}} \cap V_{j_{1}}\cap V_{j_{2}}}^{V_{j_{0}}}B_{j_{0}}  = 0,
\end{align*}
which shows that indeed $DD\B=0$. 

Now it is possible to define cohomology groups of the sheaf ${\sf F}$ on a \ipip\ $V$.
\bedefi\label{def-cohom-group}
Let  $V_{I}$ be an \ipip, endowed with the open covering of its assaying spaces, $\{V_{j}, j\in I\}$, 
and let ${\sf F}$ be the corresponding sheaf of operators on $V_{I}$. Then the $p$th cohomology group is defined as
  $ H^p(I,{\sf F}) := Z^p(I,{\sf F})/B^p(I,{\sf F}) $,  where $Z^p(I,{\sf F})$ is the set of $p$-cocycles, \emph{i.e.},   $p$-cochains $\A$ such that $D\A=0$,
 and $B^p(I,{\sf F})$ is the set of $p$-coboundaries,
 \emph{i.e.}, $p$-cochains $\B$ for which there exists a $(p-1$)-cochain $\C$ with $\B=D\C$. 
\findefi

The definition of the group $H^p(I,{\sf F})$ is motivated by the fact that the  $p$-cocycle $\A$ is given only up to a coboundary 
$\B=D\C: D\A=0=D(\A+\B)=D(\A+D\C)$.

Our definition of cohomology groups of sheaves on \ipip s depends on the particular open  covering we are choosing. 
So far we have used the open covering given by all assaying subspaces, but there might be other ones, typically consisting of unions of assaying subspaces.
 We say that an open covering $\{V_j, j\in J\}$   of an \ipip\  $V_I $ is finer than another  one,
$\{U_k, k\in K\}$, if there exists an application $t: J\to K$ such that  $V_j \subset U_{t(j)}, \forall\, j\in J$. 
For instance, $U_{t(j)}$ could be a union   $\cup_{l}V_l $   containing $V_j$.
According to the general theory of sheaf cohomology 
\cite {maclane2}, this induces group homomorphisms $t_J^K:  H^p(K,{\sf F})  \to H^p(J,{\sf F}) , \, \forall\, p\geq 0$. It is then possible to introduce cohomology groups
 that do not depend on the particular open covering, namely, by defining
$H^p(V,{\sf F})$     as the inductive limit of the groups $H^p(J,{\sf F}) $  with respect to the homomorphisms $t_J^K$. 
 
Using a famous result of Cartan and Leray 
 and an unpublished work of J. Shabani, we give now a theorem which characterizes the cohomology of sheaves on \ipip s. 
  We collect in Appendix B the definitions and results needed for this discussion.
We know that an \ipip\ $V$ endowed with the Mackey topology   is a separated (Hausdorff) locally convex space, but it is not necessarily  paracompact,
 unless $V$ is metrizable, in particular, a Banach or a \hs.   In that case, it is possible to define a fine sheaf (see Definition B.3) of operators on $V$ and apply the 
Cartan-Leray  Theorem \ref{theo:CL}. This justifies the restriction of metrizability in the following theorem.

\betheo    
Given an \ipip\ $V_{I}$, define the sheaf    ${\sf F}: V_r \to F(V_r) :=\mathrm{Op}_r =\{A_{r} : V_r \to V\}$.
Then, if $V$ is metrizable for its Mackey topology, the sheaf {\sf F} is acyclic, that is, the cohomology groups $H^p(V,{\sf F})$ are trivial for all 
 $p\geq 1$,\; \emph{i.e.}, $H^p(V,{\sf F}) = 0$.
\entheo
\prf
  First of all,  $V$ is a paracompact space, since it is   metrizable. 
  
Next, we check that the sheaf ${\sf F}: V_r \to {\sf F}(V_r) =\mathrm{Op}_r =\{A_{r} : V_r \to V\}$  on the paracompact space $V$ is fine.  
Indeed, let $\{V_{j}, j\in J \}$  be a locally finite open covering of $V$. We can associate to the latter a partition of unity  $\{\varphi_{j}, j\in J \}$. 
Then we can use $\{\varphi_{j}, j\in J \}$   to define the homomorphisms   $h_{j}: {\sf F}\to  {\sf F}$. Let $V_{r} \subset \cup_{s\in K} V_{s}$ for some index set $K$.
 For each $s\in K$, we consider the set of operators $\mathrm{Op}_s = \{A_{\w s} : V_s\to V\}$   and we define $h_j: A_{s}\mapsto h_{j}(A_{s}): =  \varphi_{j}A_{s}.$
This defines a homomorphism   $\mathrm{Op}_s  \to \mathrm{Op}_s  $ and then a homomorphism  ${\sf F}\to {\sf F}$. One can check that  
$h_j: {\sf F} \to {\sf F}$ satisfies conditions (1) and (2) in Definition B.3  and thus ${\sf F}$ is a fine sheaf. 

  Then the result follows from Theorem \ref{theo:CL}.   
\qed
\medskip

The crucial point of the proof  here is the Cartan-Leray  theorem. Let us give   a flavor of the proof of a particular case of this classical result. 
Let   $[\C]$ be an element of  $H^1(V,{\sf F}) $. We want to show that $[\C]=0$. As an equivalence class, $[\C]$  can be represented by an element
 $\C= (C_{j_{0},j_{1}}, j_{0} <j_{1})\in I$  of $C^1(I,{\sf F})$, such that  $D\C = 0$, \emph{i.e.}, 
$$
(D\C)_{j_{0},j_{1},j_{2}}  = \rho_{V_{j_{0}} \cap V_{j_{1}}\cap V_{j_{2}}}^{V_{j_{1}}\cap V_{j_{2}}} C_{j_{1},j_{2}}
 - \rho_{V_{j_{0}} \cap V_{j_{1}}\cap V_{j_{2}}}^{V_{j_{0}}\cap V_{j_{2}}} C_{j_{0},j_{2}}
+ \rho_{V_{j_{0}} \cap V_{j_{1}}\cap V_{j_{2}}}^{V_{j_{0}}\cap V_{j_{1}}} C_{j_{0},j_{1}} = 0 .
$$
Using the fact that ${\sf F}$ is a fine sheaf, we can choose the partition of unity  $\{\varphi_{i}, i\in I \}$ to define the 0-cochain  
$\E=\{E_j\},$ where $E_j = -\sum_{k\in I} \varphi_{k} (C_{jk})$. This sum is well-defined since the covering is locally finite. Applying the coboundary operator we get: 
\begin{align*}
(D\E)_{j_{0},j_{1}}  &= \rho_{V_{j_{0}} \cap V_{j_{1}}}^{V_{j_{1}}} E_{j_{1}} - \rho_{V_{j_{0}} \cap V_{j_{1}}}^{V_{j_{0}}} E_{j_{0}}
=  -\sum_{k\in I} \varphi_{k} \left(\rho_{V_{j_{0}} \cap V_{j_{1}}}^{V_{j_{1}}} C_{j_{1}k} \right)
+ \sum_{k\in I} \varphi_{k} \Big( \rho_{V_{j_{0}} \cap V_{j_{1}}}^{V_{j_{0}}} C_{j_{0}k} \Big)
\\
&=\sum_{k\in I} \varphi_{k} \Big( \rho_{V_{j_{0}} \cap V_{j_{1}}}^{V_{j_{0}}} C_{j_{0}k} 
 - \rho_{V_{j_{0}} \cap V_{j_{1}}}^{V_{j_{1}}} C_{j_{1}k} \Big)
=\sum_{k\in I} \varphi_{k} \Big(\rho_{V_{j_{0} \cap V_{j_{1}}\cap V_{k}}} ^{V_{j_{0}} \cap V_k} C_{j_{0}k}
 - \rho_{V_{j_{0} \cap V_{j_{1}}\cap V_{k}}} ^{V_{j_{1}}\cap V_k} C_{j_{1}k} \Big).
\end{align*}
Putting $k=j_{2}$  and using the equation $(D\C)_{j_{0},j_{1},j_{2}}  =0$, we find
$$
(D\E)_{j_{0},j_{1}} = \sum_{j_{2}\in I} \varphi_{j_{2}} \Big(\rho_{V_{j_{0} \cap V_{j_{1}}\cap V_{j_{2}}}} ^{V_{j_{0}} \cap V_{j_{1}}} 
C_{j_{0}j_{1}}\Big).
$$
 But since $\{\varphi_{i}, i\in I \}$  is a partition of unity, we get  $(D\E)_{j_{0},j_{1}} = C_{j_{0}j_{1}} $. 
 This means that  $D\E = \C$ and thus $[\C]=0$.  

\bigskip

Of course, the restriction that $V$ be metrizable is quite strong, but still the result applies to a significant number of interesting situations. For instance:
\bei
\vspace*{-2mm}\item [(1)] A \emph{finite} chain of reflexive Banach spaces or \hs s, for instance a triplet of \hs s
$ \H_{1}\hookrightarrow \H_{0} \hookrightarrow \H_{\ov{1}} $
or any of its refinements, as discussed in \cite[Sec.5.2.2]{pip-book}.

\vspace*{-2mm}\item [(2)] An \ipip\ $V_I$ whose extreme space $V$ is itself a \hs, like the LHS of functions analytic in a sector described in \cite[Sec.4.6.3]{pip-book}.

\vspace*{-2mm}\item [(3)] A Banach Gel'fand triple, in the sense of Feichtinger \cite{dorf-fei-gro},
 that is, a RHS (or LBS) in which the extreme spaces are (nonreflexive) Banach spaces. A nice example, extremely useful in Gabor analysis, is the so-called
 \emph{Feichtinger algebra}  $\cS_{0}(\RN^d)$, which generates the triplet 
\be\label{eq:tripletS0}
 \cS_{0}(\RN^d) \hookrightarrow L^2(\RN^d) \hookrightarrow \cS_{0}^\times(\RN^d).
\en
 The latter can often replace the familiar Schwartz triplet of tempered distributions. Of course, one can design all sorts of LHSs of LBSs interpolating between the extreme spaces, 
as explained in \cite[Sec.5.3 and 5.4]{pip-book}.
 \eni

\noindent In fact, what is really needed for the Cartan-Leray theorem is not that $V$ be metrizable, but that it be paracompact. 
Indeed, without that condition the situation  becomes totally unmanageable.
However, except for metrizable spaces, we could not find interesting examples of \ipip s with $V$ paracompact.

\subsection{ Cohomology of operator cosheaves} 
\label{sec-cohomolcosheaf}

 It is also possible to get cohomological concepts on cosheaves defined from \ipip s. The assaying spaces can also be considered as closed sets. 
Let thus $\{W_j, j\in I\}$   be such a closed covering of $V$ and let ${\sf G}$ be a cosheaf of operators defined in \eqref{cosheafunivmult}, namely,
${\sf G}: W_j \mapsto L \mathrm{Op}_j =\{A\upharpoonright W_j : $ $A \in L \mathrm{Op}(V_I) \}$  
together with maps  $\delta ^{W_{j}}_{W_{i}}: {\sf G}(W_{i})\to {\sf G}(W_{j}),\, W_{i} \subset  W_{j}$  as above.
In the sequel,  ${\sf G}(W_{j_{0}}\cup W_{j_{1}} \cup\ldots \cup W_{j_{p}}) $ denotes the set of operators 
 $W_{j_{0}}\cup W_{j_{1}} \cup\ldots \cup W_{j_{p}}\to V$. 
   Alternatively, one could consider the cosheaf defined in \eqref{eq:opcosheaf},
$\widehat{\sf G}: W_j \mapsto  \mathrm{\widehat {Op}}_j   :=\{A_{j\w} : A \in \mathrm{Op}(V_I), j\in {\sf i}(A) \}$,
and proceed in the same way.

In  this setup, we may introduce the necessary cohomological concepts, as in Definitions \ref{def_cochain} and \ref{def_cobdy}.
\bedefi
A \emph{$p$-cochain} with values in the cosheaf ${\sf G}$  is a map which associates to each union of closed sets  
$W_{j_{0}}\cup W_{j_{1}} \cup\ldots \cup W_{j_{p}}, \, j_{k}\in I, \,j_{0} < j_{1}< \ldots< j_{p} $,  
  an operator   $A_{j_{0},j_{1},\ldots, j_{p}}$ of ${\sf G}(W_{j_{0}} \cup W_{j_{1}}\cup \ldots\cup W_{j_{p}})$.
 A $p$-cochain is thus a set  $\gA = \{A_{j_{0},j_{1},\ldots, j_{p}},  \,j_{0} < j_{1}< \ldots< j_{p}\}$. 
The set of such $p$-cochains   {will be denoted by $\widehat{C}^p(I,{\sf G})$.}
\findefi
\bedefi
One can then introduce the   coboundary operator $\widehat D$ as follows:
\begin{align*}
 &\widehat D: \widehat{C}^p(I,{\sf G}) \to \widehat{C}^{p+1}(I,{\sf G}): \gA =\{A_{j_{0},j_{1},\ldots, j_{p}}\}\mapsto \{(\widehat{D}\gA)_{j_{0},j_{1},\ldots, j_{p},j_{p+1}}\},
\\
&(\widehat{D}\gA)_{j_{0},j_{1},\ldots, j_{p},j_{p+1}}= 
\sum_{k=0}^{p+1}(-1)^{k}\,\delta^{W_{j_{0}} \cup W_{j_{1}}\cup \ldots\cup W_{j_{p+1}}}_{W_{j_{0}} \cup W_{j_{1}}\cup\ldots 
\cup \widehat W_{j_{k}}\cup\ldots\cap W_{j_{p+1}}}\, A_{j_{0},j_{1},\ldots, \widehat j_{k},\ldots,  j_{p+1}}\, .
\end{align*}

 \findefi
In view of the properties of the maps  $\delta^{\w}_{\w}$, we check that $\widehat{D}\widehat{D}=0$.  

In the case of 0-cochains, a straighforward application of the formula leads to  
$$
(\widehat{D}\gA)_{j_{0},j_{1}}  = \delta^{W_{j_{0}} \cup W_{j_{1}}}_{W_{j_{1}}} A_{j_{1}} - \delta^{W_{j_{0}} \cup W_{j_{1}}}_{W_{j_{0}}} A_{j_{0}}.
$$
And if we put this to zero, we get the  constraint  
 $\delta^{W_{j_{0}} \cup W_{j_{1}}}_{W_{j_{1}}} A_{j_{1}} = \delta^{W_{j_{0}} \cup W_{j_{1}}}_{W_{j_{0}}} A_{j_{0}}$,
 which has to be satisfied in order to build a cosheaf, according to condition (CS$2$). On 1-cochains, the coboundary action gives
$$
(\widehat{D}\gA)_{j_{0},j_{1},j_{2}}  = \delta^{W_{j_{0}} \cup W_{j_{1}}\cup W_{j_{2}}}_{W_{j_{1}}\cup W_{j_{2}}} A_{j_{1},j_{2}}
 - \delta^{W_{j_{0}} \cup W_{j_{1}}\cup W_{j_{2}}}_{W_{j_{0}}\cup W_{j_{2}}} A_{j_{0},j_{2}}
+ \delta^{W_{j_{0}} \cup W_{j_{1}}\cup W_{j_{2}}}_{W_{j_{0}}\cup W_{j_{1}}} A_{j_{0},j_{1}}\, .
$$
The cohomology groups of the cosheaf   $ \widehat{H}^p(I,{\sf G}) := \widehat{Z}^p(I,{\sf G})/\widehat{B}^p(I,{\sf G}) $, with obvious notations, can then be defined in a natural way. 
Similarly for $\widehat{H}^p(V,{\sf G})$ and $\widehat{H}^p(I,\widehat{\sf G}), \widehat{H}^p(V,\widehat{\sf G})$.

Now it is tempting to proceed as in the case of sheaves and define the analog  of a fine cosheaf, in such a way that one can apply  a result similar to the Cartan-Leray theorem.
But this is largely unexplored territory, so we won't venture into it.

\section{Outcome}
\label{sec-outcome}

 The analysis so far shows that several aspects of the theory of \ipip s and operators on them have a natural formulation in categorical terms.
Of course, this is only a first step, many questions remain open. 
  For instance, does there exist a simple characterisation of the dual category  ${\bf PIP}^{\,\rm op}$ ?
Could it be somehow linked to the category of partial *algebras, in the same way as
 ${\bf Set}^{\,\rm op} $ is isomorphic to the category of complete atomic Boolean algebras 
(this is the so-called Lindenbaum-Tarski duality \cite[Sec. VI.4.6]{johnstone})?

  In addition, our constructions yield new concrete examples of sheaves and cosheaves, namely, (co)\-sheaves of operators on an \ipip, and this is probably the most important result
of this paper. Then, another open question concerns the cosheaf  cohomology groups. Can one find conditions under which the cosheaf is acyclic, that is, $\widehat{H}^p(V,{\sf G}) = 0$, 
for all  $p\geq 1$, or, similarly, $\widehat{H}^p(V,\widehat{\sf G}) = 0$, for all  $p\geq 1$?

  In guise of conclusion, let us note that cosheaf is a new concept which was introduced in a logical framework in order to dualize the sheaf concept  \cite{lambert}.
In fact one knows that the category of sheaves (which is in fact a topos) is related to Intuitionistic logic and Heyting algebras,
in the same way as the category of sets has deep relations with the classical proposition logic and Boolean algebras \cite[Sec. I.1.10]{johnstone}. 

 More precisely, classical logic satisfies the noncontradiction principle {\sf NCP}  $\neg (p \land\neg p)$) and the excluded middle principle 
 {\sf EMP}  ($p\lor \neg p$).\footnote{\emph{i.e.}, {\sf NCP} := Not-($p$ And Not-$p$) and {\sf EMP} := $p$ Or Not-$p$.}
 Intuitionistic logic satisfies {\sf NCP}, but not {\sf EMP}. Finally, we know that Paraconsistent logic, satisfying {\sf EMP}
  but not {\sf NCP}, is related to Brouwer algebras, also called co-Heyting algebras \cite{james,mckinsey}. Then, it is natural  to wonder what is the category, if any, 
 (mimicking the category of sets for the classical case and the topos of sheaves for the intuitionistic case), 
corresponding to Paraconsistent logic? The category of cosheaves can be a natural candidate for this. And this is the reason why 
it was tentatively introduced in a formal logic context. The category of closed sets of a topological space happens to be a cosheaf. 
But up to now we did not know any other examples of cosheaves in other areas of mathematics. Therefore it is interesting   
to find here additional concrete examples of cosheaves in the field of functional analysis.

  In a completely different field, the search for a
 quantum gravity theory, Shahn Majid \cite{majid1, majid2}  has proposed to unify quantum field theory and general relativity using a self-duality 
principle expressed in categorical terms. His approach shows deep connections, on the one hand, between quantum concepts and 
Heyting algebras (the relations between quantum physics and Intuitionistic logic are well-known) and, on the other hand, following
 a suggestion of Lawvere \cite{majid1}, between general relativity (Riemannian geometry and uniform spaces) and co-Heyting algebras (Brouwer algebras). 
Therefore it is very interesting to shed light on concepts arising naturally from Brouwer algebras and this is precisely the case of cosheaves. 
Sheaves of operators on \pip s are connected to quantum physics. Is there any hope to connect cosheaves of operators 
on some \pip s to (pseudo-)Riemannian geometry, uniform spaces or to theories describing gravitation? This is an open question 
suggested by   Majid's idea of a self-duality principle (let us note that Classical logic is the prototype of a self-dual structure,
  self-duality being given by the de Morgan rule, which transforms {\sf NCP}  into {\sf EMP}!). 

\appendix
\section{\hspace*{-5.5mm}ppendix A: Partial inner product spaces }

 \subsection{\pip s and \ipip s}

For the convenience of the reader, we have collected here the main features of partial inner product spaces and operators on them, keeping only what is needed for reading the paper. 
Further information may be found in our review paper \cite{at-AMP}  or our monograph \cite{pip-book}.

The general framework is that of a \pip\ $V$,  corresponding to the linear compatibility $\com$, that is,  
a symmetric binary relation $f \com g$  which preserves linearity.
We   call \emph{assaying subspace} of $V$ a  subspace $S$ such that $S^{\#\#} = S$ and 
we denote by ${\F}(V,\com)$   the family of all assaying subspaces of $V$, ordered  by inclusion. 
The assaying subspaces are denoted by $V_{r}, V_{q} , \ldots $ and the index set is $F$.  By definition, $q \leq r$ if and only if $V_{q} \subseteq V_{r}$.
 Thus we may write
\be\label{eq:gener2}
f\com g \; \Leftrightarrow \; \exists \;r \in F  \mbox{ such that } f \in V_{r}, g \in V_{\overline{r}}\,.
\en

General considerations imply that the family   ${\F}(V,\com):= \{ V_r, r\in {F} \}$, ordered  by
inclusion, is  a complete involutive lattice, \emph{i.e.}, it is stable under the following operations, arbitrarily iterated:
\medskip

\begin{tabular}{lccl}
. involution: &$V_r$                     & $\!\!\!\leftrightarrow\!\!\!$& $V_{\ov{r}}=(V_r)^{\#},$\\
. infimum:   & $V_{p \wedge q}$  &$\!\!\! :=\!\!\!$                & $V_p \wedge V_q = V_p \cap V_q,$   \qquad $(p,q,r \in {F})$\\
. supremum: & $V_{p \vee q}$     &$\!\!\! :=\!\!\!$               &$ V_p \vee V_q = (V_p + V_q)^{\#\#}$.
\end{tabular}
\\[2mm]
\noi The smallest element  of $ {\F }(V,\com)$ is  $V\co = \bigcap_r V_r $ and the greatest element   is $V = \bigcup_r V_r$.  

By definition, the index set  ${F}$    is  also a complete involutive  lattice; for instance,  
 $$
 (V_{p \wedge q})\co = V_{\ov{p \wedge q}} 
=  V_{\ov {p} \vee \ov {q}} = V_{\ov {p}} \vee V_{\ov {q}}.
$$

Given a vector space $V$ equipped with a linear compatibility $\com $, a \emph{partial inner product}   on   $(V, \,{\com})$ is a
   Hermitian form  $\ip{\cdot}{\cdot}$ defined exactly on compatible pairs of vectors. 
A \emph{partial inner product space}  (\pip)  is a  vector space $V$ equipped with a linear compatibility  and a partial inner product.

From now on, we will assume that our \pip\  $(V, \com, \ip{\cdot}{\cdot})$ is \emph{nondegenerate}, 
that is,    $\ip{f}{g} = 0   $ for all $ f \in  V^{\#} $ implies $ g = 0$.  As a consequence,  $(V\co, V)$ and  
 every couple $(V_r , V_{\ov r} ), \,  r\in {F}, $  are a  dual pair in the sense of topological vector spaces \cite{kothe}. 
Next we assume that every $V_{r}$ carries  its Mackey topology $\tau(V_{r},V_{\ov{r}})$, so that its conjugate dual is $(V_r)^\times = V_{\ov {r}}, \; \forall\, r\in {F} $.
Then,   $r<s$ implies $V_r \subset V_s$, and the embedding operator $E_{sr}: V_r \to V_s$  is continuous and has dense range. In particular, $V\co$ is dense in every $V_{r}$.

As a matter of fact,   the whole structure can be reconstructed from a fairly
 small subset of $\F$, namely, a \emph{generating}    involutive sublattice $\I$   of $\F(V, \com)$, indexed by $I$, which means that
\be\label{eq:gener}
f\com g \; \Leftrightarrow \; \exists \;r \in I \mbox{ such that } f \in V_{r}, g \in V_{\overline{r}}\,.
\en
The resulting structure is called  an \emph{\ipip} and denoted simply by $V_{I} := (V, \I, \ip{\cdot}{\cdot}) $.

Then an \ipip\  $V_{I}$ is said to be:
\bei
\vspace*{-1mm}\item [(i)]
\emph{additive}, if $V_{p \vee q} = V_{p} + V_{q}, \; \forall \, p, q \in I$.

\vspace*{-2mm}\item [(ii)] \emph {projective}  if  
$V_{p \wedge q}|_{\tau} \simeq (V_{p} \cap V_{q})_{\mathrm{proj}}, \;\forall \, p, q \in I ;$
here $V_{p \wedge q}|_{\tau}$ denotes $V_{p \wedge q}$ equipped with the Mackey topology $\tau (V_{p \wedge q}, V_{\ov{p} \vee \ov{q}})$, 
the r.h.s. denotes $V_{p} \cap V_{q}$ with the topology of the projective limit from $V_{p}$ and $V_{q}$ and $\simeq $ denotes an isomorphism of locally convex
 topological spaces.

\eni
For practical applications, it is essentially sufficient to restrict oneself to the case of an \ipip\ satisfying the following conditions:
\bei
\vspace*{-2mm}\item [(i)]
 every $V_{r}, r\in I$, is a \hs\ or a reflexive Banach space, so that the Mackey topology $\tau(V_{r},V_{\ov{r}})$ coincides with the norm topology;

\vspace*{-2mm}\item [(ii)]   there is a unique self-dual, Hilbert,  assaying subspace $V_{o} =V_{\overline{o}}$.
\eni

\noi In that case, the structure  $V_{I} := (V, \I, \ip{\cdot}{\cdot}) $ is called, respectively, 
a  \emph{lattice of \hs s} (LHS)  or a  \emph{lattice of Banach spaces} (LBS) (see \cite{pip-book} for more precise definitions, including explicit requirements on norms). 
The important facts here are that 

\bei
\vspace*{-2mm}\item [(i)] Every projective \ipip\  is additive.

\vspace*{-2mm}\item [(ii)]  A  LBS or a LHS is projective if and only if it is additive.
\eni
\vspace*{-2mm}
Note that $V\co, V $ themselves usually do \emph{not} belong to the family $\{V_{r}, \,r\in I\}$, but they can be recovered  as
$$
V\co = \bigcap _{r\in I}V_{r}, \quad V =\sum_{r\in I}V_{r}.
$$ 
 A standard, albeit trivial,  example is that of a Rigged Hilbert space (RHS) $\Phi \subset \H \subset \Phi\co$
(it is trivial because the lattice $\F$ contains only three elements). One should note that the construction of a RHS 
from a directed family of \hs s,  via projective and inductive limits, 
has been investigated recently by Bellomonte and Trapani \cite{bell-trap}. Similar constructions, in the language of categories, may be found in 
the work of Mityagin and Shvarts  \cite{mityagin} and that of Semadeni and Zidenberg \cite{semadeni}.
\medskip

Let us give some concrete examples.
 \smallskip

\noi{\sl (i)   Sequence spaces }
\smallskip

Let  $V $ be  the space   $\omega$    of \emph{all} complex sequences $x = (x_n)$  and define on it (i)
a  compatibility relation by  $x {\com} y \Leftrightarrow \sum_{n=1}^\infty  |x_n \, y_n | < \infty$; (ii)
a partial inner product $ \ip{x}{y} = \sum_{n=1}^\infty  \overline{x_n} \, y_n $.
Then   $\omega ^{\#} =  \varphi $ , the space of   finite sequences, and
the complete lattice ${\F}(\omega,{\com})$ consists of K\"{o}the's perfect sequence spaces \cite[\S\,30]{kothe}.
  {Among these, a nice example is the lattice of  the so-called     $\ell_{\phi}$ spaces associated to symmetric norming functions or, more generally, Banach sequence ideals
discussed in \cite[Sec.4.3.2]{pip-book} and previously in \cite[\S\,6]{mityagin}
(in this example, the extreme spaces are, respectively, $\ell^1$ and $\ell^\infty$).}

\medskip

\noi{\sl (ii) Spaces of locally integrable functions}
\smallskip

Let  $V $ be $ L^1_{\rm loc}(\RN, \ud x)$, the space of Lebesgue measurable functions, integrable over compact subsets, and define 
a compatibility relation on it  by $f \com g\Leftrightarrow \int_{\RN} |f(x)g(x)|  \ud x < \infty$ and
a partial inner product  $ \ip{f}{g} = \int_{\RN} \overline{f(x)} g(x)   \ud x$.
Then $V^{\#} =  L^\infty_{\rm c}(\RN, \ud x)$, the space of   bounded measurable functions of compact support.
The complete lattice ${\F }(L^1_{\rm loc},{\com})$ consists of the so-called K\"{o}the function spaces. 
  {Here again, normed ideals of measurable functions in $L^1([0,1], \ud x)$ are described in \cite[\S\,8]{mityagin}.}
\medskip

\subsection{Operators on \ipip s}
\label{sec:oper}

Let $V_{I}$ and $Y_{ K}$ be two nondegenerate \ipip s (in particular, two LHSs or LBSs). Then  an \emph{operator} from $V_I$  to $Y_{ K}$ is a map
from a subset $\D (A) \subset V$ into $Y$, such that
\smallskip

(i) $\D(A) = \bigcup_{q\in {\sf d}(A)} V_q$, where ${\sf d}(A)$ is a nonempty subset of $I$;
\smallskip

(ii)  For every $r \in  {\sf d}(A )$, there exists $u\in K$ such that the restriction of $A$ to $V_{r}$ is a continuous linear map into $Y_{u}$ (we denote this restriction by $A_{ur})$;
\smallskip

(iii) $A$ has no proper extension satisfying (i) and (ii).
\medskip

\noi We denote by Op$(V_I,Y_K)$  the set of all operators from  $V_I$ to $Y_{K}$ and, in particular, $ \mathrm{Op}(V_I) : = \mathrm{Op}(V_I,V_I)$.
 The continuous linear operator $A_{ur}: V_r \to Y_{u}$ is called a \emph{representative} of $A$.
   Thus the operator $A$ may be identified with   the collection of its representatives,
$A \simeq \{ A_{ur}: V_{r} \to Y_{u} \mbox{ exists as a bounded operator} \}.$
We will also need the following sets:
\vspace*{-2mm}\begin{align*}
{\sf d}(A) &= \{ r \in I : \mbox{there is a } \,   u \; \mbox{such that}\; A_{ur} \;\mbox{exists} \},
\\
{\sf i}(A) &= \{ u \in K : \mbox{there is a } \, r \; \mbox{such that}\; A_{ur} \;\mbox{exists} \}.
\end{align*}
\\[-4mm]
The following properties are immediate:
\bei
\vspace*{-2mm}
\item [{\bf .}]   
${\sf d}(A)$ is an initial subset of $I$:  if $r \in {\sf d}(A)$ and $r' < r$, then $r \in {\sf d}(A)$, and $A_{ur'} = A_{ur}E_{rr'}$,
 where  $E_{rr'}$ is a representative of the unit operator.  

\vspace*{-2mm}\item [{\bf .}]   
${\sf i}(A)$ is a final subset of $K$: if $u \in {\sf i}(A)$ and $u' > u$, then $u' \in {\sf i}(A)$ and $A_{u'r} = E_{u'u} A_{ur}$.
\eni

\medskip

Although an operator may be identified with a separately continous  sesquilinear form on $V^\# \times V^\#$,
it is more useful to keep also the \emph{algebraic operations} on operators, namely:
 \bei
\vspace*{-1mm}\item[(i)] \emph{Adjoint:}
every $A \in\mathrm{Op}(V_I,Y_K)$ has a {unique} adjoint $A\ta \in \mathrm {Op}(Y_K,V_I)$
and one has $A\ta{}\ta = A, $ for every $ A \in {\rm Op}(V_I,Y_K)$: no extension is allowed, by the maximality condition (iii)  of the definition.
\vspace*{-1mm}
\item[(ii)] \emph{Partial multiplication:}
Let $V_I$, $W_L$, and $Y_K$  be nondegenerate \ipip s (some, or all, may coincide).
 Let $A \in   \mathrm{Op}(V_I,W_L)$ and  $B \in   \mathrm{Op}(W_L,Y_K)$. We say that the product $BA$ is defined if and only  if
there is a $t \in{\sf i}(A) \cap{\sf d}(B)$, that is, if and only if   there is continuous factorization through some $W_t$:
\be\label{eq:mult}
V_r \; \stackrel{A}{\rightarrow} \; W_t \; \stackrel{B}{\rightarrow} \; Y_u , \quad\mbox{\emph{i.e.},} \quad  (BA)_{ur} = B_{ut} A_{tr}, \,\mbox{ for some } \;
r \in{\sf d}(A) , u\in {\sf i}(B).
\en
\eni
Among operators on \ipip s, a special role is played by morphisms.

An operator     $A\in \mathrm{Op}(V_{I},Y_{K})$ is called a \emph{homomorphism} if
\bei
\vspace*{-2mm}\item [(i)] for every $r\in I$, there exists $u\in K$ such that both $A_{ur}$ and $A_{\ov u \ov r}$  exist;

\vspace*{-2mm}\item [(ii)] for every $u\in K$, there exists $r\in I$ such that both $A_{ur}$ and $A_{\ov u \ov r}$  exist.
\eni
\vspace*{-2mm}

We denote by Hom($V_{I},Y_{K}$) the set of all homomorphisms from $V_{I}$ into $Y_{K}$ and by $\mathrm{Hom}(V_{I})$ those from $V_{I}$ into itself.
The following properties are immediate.

\beprop \label{prop:homom} Let $ V_{I},Y_{K},\ldots$ be indexed  \pip s. Then:\vspace*{-1mm}
\begin{itemize}
\item [(i)]
$A\in \mathrm{Hom}(V_{I}, Y_{K})$ if and only if  $A\ta\in \mathrm{Hom}(Y_{K},V_{I})$.
\vspace*{-2mm}
\item [(ii)]   The product of any number of homomorphisms (between successive \pip s) is defined and is a homomorphism.
\vspace*{-2mm}
\item [(iii)]    If $A\in\mathrm{Hom}(V_{I}, Y_{K})$, then $f\com g$ implies  $Af\com Ag$.
\vspace*{-2mm}
 \end{itemize}
\enprop
The definition of homomorphisms just given is tailored  in such a way that one may consider the category {\bf PIP} of all  \ipip s, with the 
homomorphisms as morphisms (arrows), as we have done in Section \ref{subsec:PIP} above. 
In the same  language, we may define particular classes of morphisms, such as monomorphisms, epimorphisms and isomorphisms.
\begin{itemize}
\vspace*{-2mm} \item [(i)] Let $M \in \mathrm{Hom}(W_{L},Y_{K})$. Then $M$ is called a \emph{monomorphism} if $MA=MB$ implies  $A=B$, 
for any two elements of $A,B\in\mathrm{Hom}(V_{I}, W_{L})$, where $V_{I}$ is any \ipip.

\vspace*{-2mm} \item [(ii)] Let $N \in \mathrm{Hom}(W_{L},Y_{K})$. Then $N$ is called an \emph{epimorphism} if $AN=BN$ implies  $A=B$, for any two elements
  $A,B\in\mathrm{Hom}(Y_{K},V_{I})$, where $V_{I}$ is any \ipip.

\vspace*{-2mm} \item [(iii)] An operator $A\in \mathrm{Op}(V_{I},, Y_{K})$ is an \emph{isomorphism} if   $A \in \mathrm{Hom}(V_{I}, Y_{K})$
and there is a homomorphism $ B \in \mathrm{Hom}(Y_{K},V_{I})$ 
such that $BA =  \textit{1}_{V}, AB =  \textit{1}_{Y}$, the  identity operators on $V,Y$, respectively.
\eni
\vspace*{-2mm} Typical examples of monomorphisms are the inclusion maps resulting from the restriction of a support, for instance, the natural injection 
$M^{(\Omega)}:  L^1_{\rm loc}(\Omega, \ud x) \to  L^1_{\rm loc}(\RN, \ud x)$, where $\RN = \Omega \cup \Omega'$ is the partition of $\RN$ in two measurable subsets of nonzero measure.
More examples and further properties of morphisms may be found in  \cite[Sec.3.3]{pip-book} and in \cite{iwota}.
\medskip

Finally,  an \emph{orthogonal projection} on a nondegenerate \ipip\ $V_I$, in particular,  a LBS or a LHS,  is a homomorphism 
 $P\in \mathrm{Hom}(V_I)$ such that $P^2 = P\ta = P$.

A \spip\ $W$ of a \pip\ $V$ is defined  in \cite[Sec.3.4.2]{pip-book}  as an \emph{orthocomplemented} subspace of $V$, that is,
 a vector subspace $W$ for which there exists a vector subspace $Z \subseteq V$ such that  $V = W \oplus Z$ and
\bei
\vspace*{-2mm}\item [(i)] 
 $\{f\}\co = \{f_{W}\}\co \cap \{f_{Z}\}\co \; $ for every $f \in V$, where$f= f_{W} + f_{Z}, \, f_{W}\in W, f_{Z}\in Z$; 
\vspace*{-2mm}\item [(ii)]
 if $f\in W, g\in Z$ and  $f\# g$, then $\ip{f}{g} = 0$. 
\eni
\noi
In the same Section 3.4.2 of \cite{pip-book}, it is shown that a vector subspace $W$ of a nondegenerate \pip\ is orthocomplemented if and only if it is
\emph {topologically regular}, which means that   it satisfies the following two conditions:
\bei
\vspace*{-2mm}\item [(i)]   
for every assaying subset $V_{r}\subseteq  V$, the intersections $W_{r} = W \cap V_{r}$  and $W_{\ov{r}} = W\cap V_{\ov{r}} $ are a dual pair in $V$;

\vspace*{-2mm}\item [(ii)] 
the intrinsic Mackey topology $\tau (W_{r}, W_{\ov{r}})$ coincides with the Mackey topology  
 $\tau(V_{r}, V_{\ov{r}}) |_{W_{r}} $ induced by $V_{r}$.
\eni

Then the fundamental result, which is the analogue to the similar statement for a \hs, says that
a vector subspace $W$ of the nondegenerate \pip\ $V$ is orthocomplemented if and only if it is the range of an orthogonal projection :
$$
W = PV \mbox{ and } V = W \oplus W^\bot = PV \oplus (1-P)V.
$$ 
Clearly, this raises the question, discussed in Section \ref {subsubsec:subobjects}, of identifying the subobjects of any category consisting of \pip s.

\setcounter{section}{2}
\setcounter{theo}{0}
 \section*{ Appendix B: Fine sheaves }

We collect here some classical notions  and results used in Section \ref{sec-cohomol}.
We recall first that the support of the   continuous function $\varphi : X \to \RN$ on the topological space $X$   is the closed set 
  \mbox {supp $\varphi := \text {closure} \{x\in X : \varphi \neq 0\}$,} which is the smallest  closed set outside which  $\varphi$ is zero.    The same definition applies to a distribution.
Then we recall the standard notion of a partition of unity.
\bedefi
 Let  $U = \{U_{i}, i\in I \}$ be an open covering of the topological space $X$. A set  of real and continuous functions $\{\varphi_{i}, i\in I \}$ defined on $X$ 
is called a \emph{partition of unity} with respect to $U$ if  
\bei
\vspace*{-2mm}\item [(i)]  $\varphi_{i}(x)\geq 0, \, \forall\, x\in X$ ;
\vspace*{-2mm}\item [(ii)]   supp $\varphi_{i} \subset U_{i},  \, \forall\, i\in I$ ;
\vspace*{-2mm}\item [(iii)]   each point $x\in X$  has an open neighborhood that meets supp $\varphi$ for a finite number of $ i\in I$ only;
\vspace*{-2mm}\item [(iv)]  $\sum_{ i\in I}\varphi_{i}(x) = 1, \, \forall\, i\in I$. This sum is well-defined by (iii).
\eni
\findefi
We recall that a topological space is \emph {paracompact} if it is separated (Hausdorff) and every open covering admits a locally finite open covering that is finer
\cite[\S 6]{choquet}.
Every metrizable locally convex space is paracompact, but there are non-metrizable paracompact spaces as well.
The following result is standard.
\betheo $X$ is paracompact if and only if $X$ is a separated topological space and each open covering of $X$ admits a partition of unity.
\entheo

The main use of paracompact spaces if for the definition of a fine sheaf, to which the Cartan-Leray theorem applies (see below).
\bedefi Let ${\sf F}$ be a sheaf on a paracompact topological space $X$. One says that {\sf F} is \emph{fine} if,  for every locally finite open covering
 $U = \{U_{i}, i\in I \}$   of $X$, there exists a set   of homomorphisms  $\{h_{i}: {\sf F}\to {\sf F}, i\in I \}$ such that: 
\bei
\vspace*{-2mm}\item [(1)] For each $i\in I $,  there exists a closed set $M_{i}$  of $X$ such that  $M_{i}\subset U_{i}$   and $h_{i}({\sf F}_{x})=0$  for  $x\not\in  M_{i} $,
 where ${\sf F}_{x}$  is the stalk of the sheaf ${\sf F}$ at the point $x$ (the stalk is defined by the inductive limit
 $\stackrel{\longrightarrow}{\lim}_{\,V_{k} \ni x }{\sf F}(V_{k})$).

\vspace*{-2mm}\item [(2)] $\sum_{i\in I}h_{i} = \textit{1}$. This sum is well-defined since the covering $U$ is locally finite.  
\eni
\findefi

Then the basic result is the following standard theorem.\footnote{A good introduction to the cohomology of sheaves and to the Cartan-Leray  theorem can be found in \cite{ward}.}

\betheo[Cartan-Leray] \label{theo:CL}Let  ${\sf F}$ be a fine sheaf on a paracompact topological space $X$ Then ${\sf F}$  is \emph{acyclic}, that is, 
 the higher order sheaf cohomology groups  are trivial, $H^p(X,{\sf F}) = 0$  for all  $p\geq 1$.
\entheo

\section*{Acknowledgements}

JPA thanks Juma Shabani (UNESCO) for communicating his unpublished results.   DL thanks Bertrand Hespel (FUNDP) for   inspiring discussions. 
We all thank the anonymous referee for his constructive remarks that have definitely improved the paper.


\begin{thebibliography}{99}

\bibitem{pip1} {   J-P.~Antoine} and {   A.~Grossmann}, ``Partial inner product spaces. I. General properties,"
\textit{Journal of  Functional Analysis}, vol. {23}, no.4, pp. 369--378,  1976.

\bibitem{pip2}{   J-P.~Antoine} and {   A.~Grossmann}, `` Partial inner product spaces. II. Operators,"
\textit{Journal of  Functional Analysis}, vol. {23}, no.4, pp. 379--391,  1976.

\bibitem{pip3} {   J-P.~Antoine}, ``Partial inner product spaces. III. Compatibility relations revisited,"
\textit{Journal of  Mathematical Physics}, vol. 21, no.2, pp.  268--279, 1980.

\bibitem{pip4} {   J-P.~Antoine}, ``Partial inner product spaces. IV.  Topological considerations,"
\textit{Journal of  Mathematical Physics}, vol. 21, no.8, pp.  2067--2079, 1980.

\bibitem{pip-book}  {    J-P. Antoine} and {     C. Trapani},
\textit{Partial Inner Product Spaces: Theory and Applications}, vol. 1986 of  \textit{Lecture Notes in Mathematics}, Springer, Berlin, Germany, 2009.

\bibitem{at-AMP}  {    J-P. Antoine} and {     C. Trapani}, ``The partial inner product space method: A quick overview,"
 \textit{Advances  in Mathematical Physics}, Vol. { 2010} {Article ID 457635}; Erratum, \textit{Ibid.} {Vol. { 2011},} {Article ID 272703}.

\bibitem{iwota}   {   J-P.~Antoine} and {    C.~Trapani}, ``Some classes of operators  on  partial inner product spaces,"
in \textit{Proc. IWOTA 2010}, Operator Theory: Advances and Applications; W.Arendt \emph{et al.} (eds.), 
Birkh\"{a}user,  Basel, 2012 {(to appear)}.  

\bibitem{maclane1} { S. Mac Lane}, \textit{Categories for the Working Mathematician},  2nd ed.,  Springer, Berlin, Germany, 1997. 

\bibitem{maclane2} { S. Mac Lane} and {  I. Moerdijk},  \textit{Sheaves in Geometry and Logic. A First Introduction to Topos Theory}, Springer, Berlin, Germany, 1992.

\bibitem{neeman}  { A. Neeman,} \textit{Algebraic and Analytic Geometry}, Cambridge University Press, Cambridge, UK, 2007, pp. 349-355.

 \bibitem{dorf-fei-gro}{   M.~D\"{o}rfler}, {  H.G.~Feichtinger}  and {  K.~Gr\"{o}chenig},
``Time-frequency partitions for the Gelfand triple $(\cS_{0},L^2,\cS_{0}^\times)$,"  \textit{Mathematica Scandinavica},  vol. {98}, pp. 81--96, 2006.

\bibitem{johnstone}  {    P. Johnstone}, \textit{Stone Spaces}, Cambridge University Press, Cambridge, UK, 1992.

\bibitem{lambert} { D. Lambert} and { B. Hespel},  ``From topology of conciliation to logic of contradiction," \textit{Logique et Analyse}, CNRL, Brussels, Belgium, 2011 (to appear).

\bibitem{james}{    W. James}, Closed set sheaves and their categories, in \textit{Inconsistent Mathematics}, C.~Mortensen (ed.) Kluwer, Dordrecht,  1995, pp. 115-124.

\bibitem{mckinsey}{ J.C.C. McKinsey}  and {  A. Tarski}, ``On closed elements in closure algebras,"  \textit{Annals of Mathematics}, vol. { 47}, pp. 122--162,  1946.

\bibitem{majid1} { S. Majid,} \textit{Foundations of Quantum Group Theory}, Cambridge University Press, Cambridge, UK, 1995, p. 294.

\bibitem{majid2} { S. Majid,}  ``Quantum spacetime and physical reality", in  \textit{On Space and Time}  (S. Majid, ed.) 
Cambridge University Press, Cambridge, UK, 2008, pp. 56-140 (look at the diagram p. 122).

\bibitem{bell-trap}    { G.~Bellomonte} and {     C.~Trapani}, ``Rigged Hilbert spaces and contractive families ofHilbert spaces,"
 \textit{Monatshefte f\"ur Mathematik }, 2011 (to appear) (DOI: 10.1007/s00605-010-0249-1).

\bibitem{mityagin}  { B.S. Mityagin}  and {  A.S. Shvarts}, ``Functors in the category of Banach spaces, "
\textit{Russian Mathematicsal Surveys}, vol. { 19}, pp.  65--127,  1964.

\bibitem{semadeni}    { Z. Semadeni} and {  H. Zidenberg},  ``Inductive and inverse limits in the category of Banach spaces," 
\textit{Bulletin de l'Acad\'emie Polonaise des Sciences, S\'erie des  sciences math. astr. et phys.}, vol. { 13}, no.8,   pp. 579--583, 1965.

\bibitem{kothe} { G. K\"{o}the}, \textit{Topological Vector Spaces  I},  Springer, Berlin, Germany, 1966.

\bibitem{choquet} { G. Choquet}, \textit{Lectures on Analysis, Vol.I, Integration and Topological Vector Spaces},  Benjamin, New York and Amsterdam, Netherlands, 1969.

\bibitem{ward} { R.S. Ward} and {  R.O. Wells Jr.}, \textit{Twistor Geometry and Field Theory},  Cambridge University Press, Cambridge, UK, 1990, pp. 166-197.

\end{thebibliography}
 \end{document}